\documentclass[preprint2]{aastex}
\def\eg{e.g.,}

\def\msun{\mbox{M$_\odot$}}

\def\mpch{\mbox{$h^{-1}$ Mpc}}
\def\mhubble{\mbox{km s$^{-1}$ Mpc$^{-1}$}}

\def\nfw{\mbox{{\tiny NFW}}}
\def\ome{\mbox{$\Omega_0$}}
\def\omel{\mbox{$\Omega_\Lambda$}}

\def\sige{\mbox{$\sigma_8$}}
\def\LCDM{{$\Lambda$CDM}}

\shorttitle{Evolution of NGC~6822}
\shortauthors{Carigi et al.}

\received{}
\begin{document}

\title{Chemical and Photometric Evolution of 
the Local Group Galaxy NGC~6822 in a
Cosmological Context
}

\author{Leticia Carigi}
\affil{Instituto de Astronom\'{\i}a, UNAM, 
Apdo. Postal 70-264, M\'exico 04510 D.F., Mexico}
\email{carigi@astroscu.unam.mx}
   
\author{Pedro Col\'{\i}n}
\affil{Centro de Radioastronom\'{\i}a  y Astrof\'{\i}sica, UNAM, 
Apdo. Postal 72-3, 58089 Morelia, Michoac\'an, Mexico}
\email{p.colin@astrosmo.unam.mx}

\and 

\author{Manuel Peimbert}
\affil {Instituto de Astronom\'{\i}a, UNAM,
Apdo. Postal 70-264, M\'exico 04510 D.F., Mexico} 
\email{peimbert@astroscu.unam.mx}

\begin{abstract}

Based on the photometric properties of NGC 6822 we derive a robust star formation history.
Adopting this history we compute 15 models of galactic chemical evolution.
All of them match present-day photometric properties. 
The dark halo mass in all models evolves according to the mass assembly 
history predicted by a $\Lambda$CDM cosmology. We model the evolution of the 
baryonic mass aggregation history in this cosmological context assuming 
that part  of the gas available for accretion never falls into the system
due to two different physical processes. 
For seven models we assume that during accretion  the universal baryon fraction is reduced by reionization
only. The best model of this first group, 
a complex model with an early outflow, 
fits the observed
gaseous mass, and the O/H, C/O, and Fe/O present-day values.
This model requires a lower upper mass limit for the IMF than that 
of the solar vicinity, in agreement with recent 
results by other authors. 
We have also computed eight models where, in addition to reionization,
the accreted baryon fraction is reduced by large-scale shock heating.
The best model of this series, that also requires and early outflow and
a lower upper mass limit for the IMF,
can marginally fit  the gaseous mass and 
the O/H and C/O observed values.

\end{abstract}

\keywords{galaxies: abundances---galaxies: evolution---galaxies: individual
NGC~6822---galaxies: irregular---Local Group}

\section{Introduction}  

The main aim of this paper is to produce chemical evolution models of the 
irregular galaxy NGC 6822. These models will be used to look for consistency 
with the cosmological context of galaxy formation as well as to study the 
possible presence of outflows from this object to the intergalactic medium.

In the field of chemical evolution often it is argued that
a model for a given galaxy is far from being unique. In this paper we use
as many observational constraints as possible to try to obtain a
robust model with the idea of getting closer to a unique solution. The
constraints used include: gas mass, chemical abundances, integrated colors, and a star
formation history derived from observations. 
On the other hand the models used 
a gas infall history
derived from the hierarchical standard $\Lambda$CDM cosmology.
The models also used the IMF and the yields that reproduce the chemical evolution of the
solar neighborhood and the Galactic disk \citep{car05}.

NGC~6822 is particularly suited for chemical evolution
models because its star
formation history is well known \citep{wyd01,wyd03}, its present chemical
composition might permit to decide if outflows to the intergalactic medium
have been produced by this galaxy. NGC~6822, unlike the Magellanic Clouds,
has not lost a significant amount of mass due to tidal effects.

NGC~6822 is a member of the local group that apparently has not
been affected by tidal effects from the Milky Way or M31 \citep{saw05}.
There has been no recent interaction between NGC 6822 and these two galaxies
because:
i) NGC 6822 is located 495 kpc away from our Galaxy 
and is receding from it at a radial velocity of
44 km s$^{-1}$ \citep[e.g.][]{tri00},
and
ii) NGC 6822 is also located at 880 kpc from M31,
it is separated from M31 by more than 90 degrees in the sky, and M31 is
approaching our Galaxy at 121 km s$^{-1}$ \citep{tri00}.

An open problem in the study of the chemical evolution of
irregular galaxies is the cause of the difference between the observed
effective oxygen yield and the oxygen yield predicted by closed box chemical
evolution models. One solution to this problem might
be due to the presence of O-rich galactic outflows \citep[e.g.][]{gar02}.
Several lines of reasoning indicate that O-rich outflows produced by gas rich
irregular galaxies are unlikely. Carigi et al. (1995, 1999), based on the C/O
and O/H values for nearby irregular galaxies, have found that O-rich outflows
have not played an important role in their evolution. A similar result
has been obtained by \citet{lar01}, based on the N/O versus O/H relationship.
These results indicate that O-rich outflows do not play a major role in the
evolution of irregular galaxies.
Another solution to the  O yield problem
might be due to the presence of  dark matter
\citep[e.g.][]{car99}.

The importance of outflows to the intergalactic medium from nearby
irregular galaxies depends on many factors, like their total mass,
the distribution in time and space of their star formation,
and the presence of tidal effects \citep[e.g.][and references
therein]{leg01, ten03, mar03, fra04, fuj04}.

In most galaxy chemical evolution models the 
infall of gas of primordial composition is modeled assuming
a parametric form 
\citep[\eg][]{lm03}. Here we
will take a cosmological approach: the rate at which gas is
accreted by the galaxy will depend on how the mass, dark and
baryonic, falls into the galaxy (the mass assembly history, MAH), and the
reionization history of the universe. The MAH in turn
depends on the assumed cosmology and the total mass of the galaxy
\citep[\eg][]{vladimir98, van02}. For a halo of a given mass,
there is an ensamble of possible MAHs. 
In this paper we will
take the mean MAH (defined in section 3.3) as representative of the MAH of NGC~6822.
Before the epoch of reionization, the history of the infalling 
gas is simply related by a constant to the MAH, afterwards the
amount of gas accreted by the galaxy will depend on how fast
the mass of the dark halo and the filtering mass grow with time
\citep{g00, kgk04}. 
We will  assume also that
the gas infalling history is not altered by processes such
as ram pressure or tidal striping. These assumptions are 
supported by the absence of warm-hot gas associated to 
the Local Group and by the fact that this galaxy is isolated.

In section 2 we define  different masses  present in an irregular
galaxy and their role in chemical evolution models.
The total mass and the halo mass assembly history within a $\Lambda$CDM 
cosmology are derived in sections 3.1 -- 3.3. 
In section 3.4 we discuss two
scenarios for the baryonic mass building of the galaxy, one in which
the reionization of the universe reduces the galaxy baryon fraction,
below the universal one; and another in which in addition to 
reionization the galaxy baryon fraction is also reduced by large-scale shock heating.
In section 4 we estimate the gaseous mass and discuss
the present day chemical composition to be fitted by the models. 
In section 5 we determine the variation of the star formation rate and the luminosity
with time. 
In section 6 we discuss the evidence in favor  of gas outflows.
In section 7 we present the general characteristics of the chemical
evolution models. 
In section 7.1 we discuss the large infall models, hereinafter L models, 
where the effect of  gas shock heating is not considered. 
In section 7.2 we discuss the small infall models, hereinafter S models,
where the effect of gas shock heating is considered, this effect prevents a substantial fraction
of baryons from reaching the galaxy.
In sections 8  and 9 we present the discussion and the conclusions.

\section {Definitions} 

We can define the total mass, $M_{total}$, as

\begin{equation}
M_{total}=M_{bar}+M_{DM},
\end{equation}

\noindent
where 
$M_{bar}$ is the baryonic mass, and $M_{DM}$ is the non-baryonic mass.

$M_{bar}$ can be expressed as

\begin{equation}
M_{bar} = M_{sub} + M_{stars} + M_{rem} + M_{gas},
\end{equation}

\noindent
where $M_{sub}$ is the mass in substellar objects (with stellar mass, $m$, lower than 0.1 \msun),
and has also been called the baryonic dark matter,
$M_{stars}$ is the mass of objects with $m > 0.1$ \msun,  
and
$M_{rem}$ is the mass in compact stellar remnants.

$M_{gas}$ can be approximated by

\begin{equation}
M_{gas} = M({\rm{H~I}}) + M({\rm{He~I}}) + M({\rm{H_2}}),
\end{equation}

\noindent
where we have neglected the ionized and the heavy element components.
$M_{gas}$ is also given by

\begin{eqnarray}
\nonumber
M_{gas} & = & M_{bar}({\rm{accreted}}) - M_{gas}({\rm{outflow}}) - \\
& & (M_{sub} + M_{stars} + M_{rem}) + M_R,
\end{eqnarray}

\noindent
where $M_{bar}$(accreted) is the baryonic mass accreted from the intergalactic
medium during the galaxy formation, $M_{gas}$(outflow) is the baryonic mass
ejected to the intergalactic medium during the galactic evolution, and $M_R$
is the mass returned to the ISM by the evolution of the stars.

We define the ratio of the baryonic mass accreted by the galaxy 
to the total mass as

\begin{eqnarray}
f_{acc}  = & M_{bar}({\rm{accreted}})/M_{total}, 
\end{eqnarray}

\noindent
and the ratio of the baryonic mass that remains in the galaxy
to the total mass as

\begin{eqnarray}
f_{gal}  = f_{acc} - M_{gas}({\rm{outflow}})/M_{total}.
\end{eqnarray}

In this paper, galaxy and halo are synonymous.

\section{ Total Mass: $M_{total}$}

\subsection{$M_{total}$ from direct observations of the rotation curve}  

A high-resolution rotation curve that extends to 5 kpc 
has been determined for this galaxy by \citet{wel03}
using the Australian Telescope Compact Array. 
Inside this radius, $R$, the mass of the galaxy given by
$M_{lower}= \frac{V_c^2 R}{G}$ is $3.5 \times 10^9$ \msun, where we are assuming
that the circular velocity, $V_c$, is equal to
the {\it rotational}
velocity determined observationally from the gaseous component
of the disk and amounts to 55 km s$^{-1}$. With an observed baryonic
mass of $4.3 \times 10^8$ \msun (see sections 4 and 7)
we find that NGC~6822 is a galaxy dominated by dark matter
with $M_{bar}/M_{lower} = 0.12$, inside $R< 5$ kpc.

\subsection{$M_{total}$ from the cosmological context} 

The total mass of a galaxy is a model dependent quantity. 
Here we will
use a cosmologically motivated model to estimate the mass
of the galaxy in question. Cosmological collisionless
N-body simulations 
show that density profiles of dark matter (DM) halos can be 
described well by the NFW profile \citep{NFW95,NFW96, NFW97}
\begin{equation}
\rho_{\nfw} (r) = \frac{\rho_0}{r/r_s(1 + r/r_s)^2}.
\end{equation}
The characteristic radius $r_s$ is the radius where the logarithmic
derivative of $\rho_{\nfw}$ is $-2$, while the characteristic density
$\rho_0$ is given by $\rho_0 = 4 \rho_{\nfw}(r_s)$. The NFW profile
can also be characterized by a concentration parameter, 
$c_{vir} \equiv R_{vir}/r_s$, and by the virial circular velocity
$V_{vir} = \sqrt{G M_{total}/R_{vir}}$, where $M_{total}$ is the virial
mass of the galaxy. 
The virial radius, $R_{vir}$, can be
defined as the radius where the average halo
density is $\delta$ times the mean density of the universe
according to the
spherical top-hat model. Here $\delta$ is a number that depends on
epoch and cosmological parameters (\ome,\omel); for a flat (0.3,0.7) 
\LCDM\ model $\delta \sim 334$ at the present epoch.
In practice, we use cosmology 
to infer a relationship
between the maximum circular velocity, $V_{peak}$, and $V_{vir}$, that
is given by
\begin{equation}
V_{vir}^2=  4.63 \frac{f(c_{vir})}{c_{vir}} V_{peak}^2
\end{equation}
\citep{Bullock2001}, where $f(c_{vir}) = \ln(1+c_{vir}) - c_{vir}/(1+c_{vir})$.
We thus can estimate the total mass of the halo 
once $V_{peak}$ and $c_{vir}$ are given. 

We use as $V_{peak}$
the rotational velocity at the outermost measured point. The mass
of the halo may be underestimated because: 
i) the rotation curve for this galaxy may still rise beyond 
this radius and
ii) we are neglecting any 
contribution from the gas
velocity dispersion.
The concentration of a DM halo
depends on many parameters such as nature of the
dark matter, halo mass, epoch, normalization 
of the power spectrum, etc. Here we will assume the
``concordance'' \LCDM\ model ($\ome = 0.3$, $\omel = 0.7$,
$h =0.7$, $\sige = 0.9$) where $h$ is the Hubble constant 
in units of $100 \ \mhubble$ and \sige\ is 
the rms of the mass fluctuations computed with
the top-hat window of radius $8 \ \mpch$. If we take the
median concentration estimated using the model by 
\citet{Bullock2001} and $V_{peak} = 55$ km s$^{-1}$ from \citet{wel03}
we find $M_{total} = 2.6 \times 10^{10}$ \msun.
This is the total mass that we will assign to NGC~6822
that implies an $R_{vir}=76.3$ kpc and $M_{lower}/M_{total}=0.13$.

The adopted $M_{total}$ implies
that baryons at present contribute to
the galaxy mass budget with only 1.65\% , such
small values seem to be typical of dwarf galaxies.

\subsection{Mass assembly history: MAH}

Within the CDM galaxy formation paradigm a halo grows in 
a hierarchical fashion: self-bound structures form from 
bottom to top through the accretion and merging of
smaller structures. A halo of a given mass today can 
arise from one of many possible mass aggregation histories. 
We use the recipe described in
appendix A of \citet{van02}, which is based on
the extended Press-Schechter formalism \citep{bcek91, bower91, lc93},
to derive the mean MAH for our given particular cosmological
model and halo mass.

\citet{van02} finds, after experimenting with a variety of fitting
functions, that the MAHs are well fitted by the following parametric
form:
\begin{equation}
\log \left[ \frac{M(z)}{M_0} \right] = -0.301 \left[ \frac{\log(1+z)}
{\log(1+z_f)} \right]^\nu,
\end{equation}
where $M(z)$ and $M_0$ are the mass of the halo at $z$ and $z= 0$, 
respectively, and $z_f$ and $\nu$ are free parameters. According
to this expression, $z_f$ is defined as the redshift where 
half of the mass of the halo is acquired, and it is 1.50 for the
total mass and cosmological context assumed in this paper.
Based on Figure 6 of
van den Bosch we took two additional models with 
extreme values of $z_f$, given by $\log (1 + z_f) = 0.4 \pm 0.1$, 
to cover the scatter in the mass assembly histories.  
The mean MAH is computed assuming $z_f = 1.50$, the
slow and fast MAHs are those computed adopting $z_f=1.00$ and $z_f = 2.16$,
respectively.
Figure 1 shows the MAHs of these models (solid for mean and dotted
lines for slow and fast).

\begin{figure}[ht]
\plotone{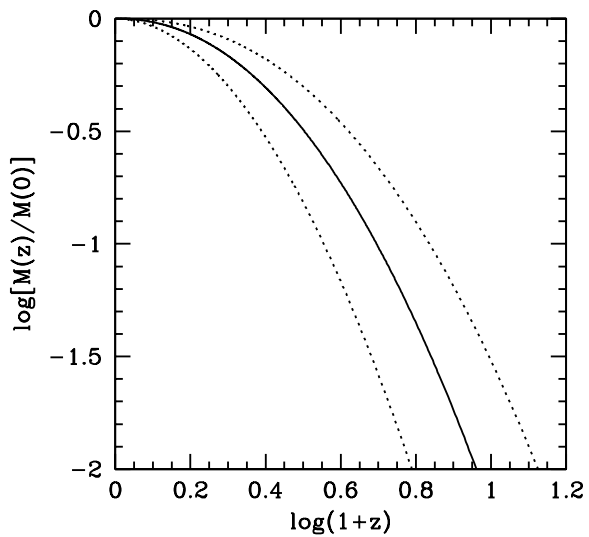}
\caption{The mean mass assembly history in units of the present-day
mass (solid line) computed using
the recipe described in appendix A of \citet{van02} with $z_f =
1.50$ (the redshift where the halo acquires half of its mass).
The scatter in the diagram has been incorporated
by using two extreme values for $z_f$ namely 1.00 and 2.16
(dotted lines). 
}
\end{figure}

\subsection{$M_{gas}$ {\it versus} time}

A high fraction of baryons $\sim 90$\% in the present-day
universe do not reside in galaxies, the missing baryons 
\citep{fp04}. They can
be in virialized regions of halos (hot gas in hydrostatic
equilibrium), in a diffuse phase with 
overdensities $\delta < 1000$ ($\delta \equiv \rho/\bar{\rho} 
- 1$, where $\bar{\rho}$ is the mean density of the universe),
or in a warm-hot phase with $10^5 < T{\rm(K)} < 10^7$
\citep{dave01, kang05, mo05}. This gas could have been 
expelled out of halos by strong winds or never became
bound to halos in the first place.

Photoionization of the IGM by quasars or early star formation is an
efficient mechanism that impedes gas to fall into small galaxies. 
Below we estimate the effect that the reionization of the universe
could have had on the gas infalling history of NGC~6822.

We use the formulas B1-B3 of appendix B of \citep{kgk04} to estimate
the effect of reionization on the structure of NGC~6822. 
The formula B3 given below
\begin{equation}
f_g(M,z) = \frac{f_b}{[1 + 0.26 M_F(z)/M]^3},
\end{equation}
is an expression for the fraction of cold gas left in the 
halo of mass $M$ \citep{g00} and
$f_b$ is the universal baryon fraction and amounts to 0.14. 
The filtering mass, $M_F$, defined
as the mass of a halo which would retain 50\% of its gas mass,
is an increasing function of cosmic time and depends on 
when reionization starts, $z_r$ (see, for example, Fig. B10 of
Kravtsov et al.). We use the mass assembly history $M(z)$ 
of section 3.3 and
equation (10) to compute the fraction of baryonic matter accreted by the galaxy
at any given redshift, that is given by
\begin{eqnarray}
\nonumber
f_{acc}(z)  =  \frac{M(z_r)}{M(z)}f_b & &    + \\
  \frac{1}{M(z)} \int_{z_r}^{z} f_g(M,z)M^\prime (z) dz, 
\end{eqnarray}
where the prime denotes derivative with respect to redshift.

In Figure 2 we plot $f_{acc}$ as a function of time for 
the mean MAH (solid line). 
The formula above is formally valid for $z < z_r$, but it is 
clear that for redshifts
greater than $z_r = 7$ this fraction is just given by $f_b$. Once
the epoch of reionization begins the amount of baryons in the
galaxy at any time will depend on how $M_F(z)/M$ evolves. 
$M_F(z)/M$ varies by about a factor 2 from $z = z_r$ ($t=0.75$ Gyr) to
$z = 0$ ($t=13.5$ Gyr, the present time) for the  MAH model. 
Its rapid increase from $z_r$ to $z \sim 4$ ($t \sim 5$ Gyr)
makes $f_{acc}$ decrease after $z_r$. As time goes on, the drop
in $f_{acc}$ is stopped because $M_F(z)/M$ decreases.

Yet, there is a second mechanism that might
reduce the amount of baryons in galaxies: assume that a 
significant fraction of baryons was never incorporated 
into the galaxy. Primordial gas could have been shock-heated
by the collapse of non-linear structures of mild overdensities 
(pancakes and filaments), raising the entropy floor \citep[\eg][]{mo05}
and thus increasing the cooling time above the Hubble time. 
Although simulations with higher numerical 
resolution are needed to confirm this prediction
\citep{mo05}, present-day simulations seem to 
support this idea. Numerical simulations by \citet{kang05} show that about
40\% of the gas at present is in a warm-hot phase (25\% with
$10^5 < T{\rm(K)} < 10^7 $ and 15\% with $T{\rm(K)} < 10^5$). We will model this
phenomena by simply assuming that this increase in temperature in the
intergalactic medium occurs suddenly, when the mass of the gas in the galaxy 
reaches 60\% of its present-day value (in the absence of large-scale 
shock heating). From this epoch, $z = 1.10$ ($t=5.37$ Gyr), to the present 
the amount of accreted baryons in the galaxy does not increase. 
The second drop in $f_{acc}$ (see Figure 2, dotted line) is due to this effect.

\begin{figure}[htb!]
\plotone{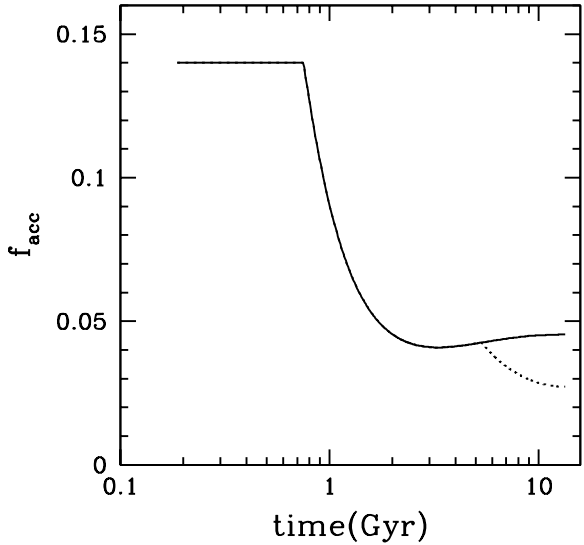}
\caption{
Evolution of the ratio of the baryonic matter accreted relative to 
the total mass for the  mean MAH,
 taking into account the effect
of reionization (solid line), and in addition to reionization the
effect of large-scale gas shock
heating after $t = 5.37$ Gyr (dotted line). 
Before $t=5.37$ Gyr the solid line represents both types of models since
the effect of large scale shock heating occurs at $t=5.37$ Gyr
and does not affect the MAH for earlier times.
We have assumed the onset of reionization 
at $t_r = 0.75$ Gyr ($z_r = 7$). 
Note that $f_{acc}$ equals the universal baryon fraction $f_b$
for $t < t_r$ and drops quickly below $f_b$ soon after and reaches a minimum.
At $t = 5.37$ Gyr ($z_r = 1.10$), which is the
time when the mass gas reaches 60\% of the total accreted mass of the large infall models, 
the curve divides in two: the solid and dotted lines. 
}
\end{figure}

\begin{figure}[htb!]
\plotone{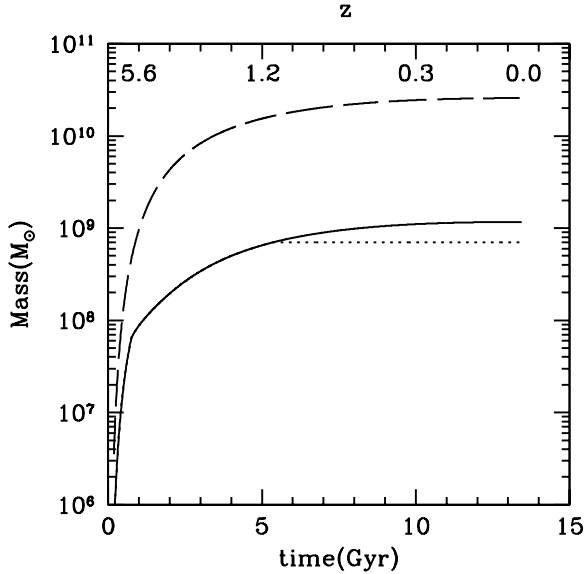}
\caption{
Dark (dashed line) and baryonic (solid and dotted lines)
mass evolution. The infall of gas follows that of the dark
matter until the onset of the reionization of the universe.
{From} then on, the gas mass growth is reduced in relation to
that of the DM. 
After $t=5.37$ Gyr in the small infall models, in which the intergalactic 
gas temperature increases due to large-scale shock heating  (dotted line), 
the galaxy stops accreting gas
once the gas mass reaches 60\% of the total accreted mass of the large
infall models (see main text); i. e., in the small infall models we assume that 40\% of the
gas will never be incorporated into the galaxy.}
\end{figure}
 
It should be clear by now that the evolution of $f_{acc}$
as well as its present-day value depend on the MAH.
For the MAHs considered for NGC~6822 and
in the absence of large-scale shock heating, 
$f_{acc}(z=0)$ takes the following values:
0.033 (slow, $z_f=1.00$), 0.045 (mean, $z_f=1.50$), and 0.066 (fast, $z_f=2.16$).
For completeness we also show in Figure 3 the evolution 
of the total (dashed line) and baryonic (solid line) mass 
for the mean MAH model, time is cosmic time. 
As in Figure 2, the dotted line represents the model where the 
large-scale shock heating of the gas is incorporated.

At this time it may be convenient to summarize the steps we have taken 
to produce the gas mass accretion history. The infall of gas is proportional
to that of the dark matter {\it until the onset} of the reionization epoch which 
we have taken to be $z=7$. Once the reionization starts the accretion rate of gas 
is no longer proportional to that of the dark matter. In fact, in some models, models 
with shock heating, the fall of gas is stopped while the accretion of dark 
matter continues. In this sense, the baryonic buildup of the galaxy is not
certainly hierarchically. We will see later that the model favored by the 
observed chemical abundances requires an early galactic wind which in turn
makes even less hierarchical the baryonic buildup of NGC~6822. 
This non hierarchical scenario has been called the downsizing problem
and has been discussed elsewhere (e.g. Arnouts et al. 2005, Cowie et al. 1996).

In section 7 we will describe two families of models
for the increase of $M_{gas}$ {\it versus} time. Those models
that accrete a relatively large $M_{gas}$ and consider the
effects of reionization only, will be called
{\it large infall models} (L models), and their chemical evolution will be
discussed in section 7.1. Those models that accrete a relatively
small $M_{gas}$ and that in addition to reionization
consider the effect of shock heating will be
called {\it small infall models} (S models) and will be discussed in section 7.2.

\section{Present Mass and Chemical Composition of the Gaseous Component}  

\subsection{Chemical composition} 

As observational constraints of the evolution models we will use the
present chemical composition of the interstellar medium of NGC 6822.
To this effect we will combine the abundances of the supergiant stars and the
\ion{H}{2} regions. The supergiant stars were formed a few million years ago, 
therefore we expect them to have the same abundances than the \ion{H}{2} regions.

In Table~\ref{tta} we present the adopted NGC 6822 abundances. 
With the exception of the Fe abundance all the abundances for NGC 6822 are those derived 
by \citet{pei05} for Hubble V, the brightest \ion{H}{2} region of the galaxy.
The Fe abundance was derived by
\citet{ven01} from two A supergiants and is in excellent
agreement with the value derived by \citet{mus99} from three B supergiants.
The gaseous Fe abundance derived from the \ion{H}{2} regions can not be used
as a constraint because most of the Fe atoms are trapped in dust grains.
The O abundance is in very good agreement with the O/H values derived
by \citet{ven01} for the two A supergiants mentioned before that amount to 12 + log O/H = 8.36.

There has been some discussion about the possible presence of abundance gradients in 
irregular galaxies, for example Pagel et al. (1978) state that a spatial
abundance  gradient, an analogous to those found in spiral galaxies, 
is small or absent in the LMC and conspicuously absent in the SMC.
The accuracy of the data on \ion{H}{2} regions and supergiant stars 
for NGC 6822 is not yet good enough to establish the
presence of a gradient. Works by \citet{vm02} and \citet{lee05} are not
conclusive, but seem to indicate 
that the O/H gradient is small or absent.
NGC 6822 V is located about 1 kpc away from the galactic center
and according to Gottesman \& Weliachew  (1977) about 70 \% of
the gaseous mass ($M({\rm HI})=1\times 10^8$ \msun)
is located within 2.6 kpc.
In what follows we will consider that the chemical composition of NGC 6822 V
is representative of the gaseous component of NGC 6822.

Also in Table~\ref{tta} we present the Orion and solar abundances for
comparison. For Orion: Fe comes from B stars \citep{cun94}
and the other elements from the \ion{H}{2} region \citep{est04}. 
The solar abundances come from \citet{ags05},
with the exception of the He abundance that corresponds to the initial value
computed by \citet{chr98} .

The Orion nebula is more representative of the present
chemical composition of the solar vicinity than the Sun for two main
reasons: i) the Sun appears deficient by roughly 0.1 dex in O, Si, Ca, Sc, Ti,
Y, Ce, Nd and Eu, compared with its immediate neighbors with similar iron
abundances \citep{all04}, the probable reason for this difference is that the Sun is
somewhat older than the comparison stars, and
ii) all the chemical evolution models of the Galaxy predict a
steady increase of the O/H ratio in the solar vicinity with time,
for example the chemical evolution model of the solar vicinity presented by 
\citet{car03}, \citet{car05}, and \citet{ake04} 
indicates that the O/H value in the solar vicinity has
increased by 0.13~dex since the Sun was formed.

\begin{deluxetable}{lrrr}
\tablecaption{NGC~6822,  Orion, and Solar Total Abundances
\label{tta}}
\tablewidth{0pt}
\tablehead{
\colhead{Element}  &
\colhead{NGC~6822\tablenotemark{a}} &
\colhead{Orion\tablenotemark{b}} &
\colhead{Sun\tablenotemark{c}}}
\startdata
12 + log O/H         & $8.42 \pm 0.06 $ & $8.73 \pm 0.03 $   & $8.66 \pm 0.05$   \\
log C/O              & $-0.31 \pm 0.13 $ & $-0.21 \pm 0.04 $   & $-0.27 \pm 0.10 $  \\
log N/O              & $-1.37 \pm 0.17 $ & $-1.00 \pm 0.10 $   & $-0.88 \pm 0.12 $  \\
log Fe/O             & $-1.41 \pm 0.10 $ & $-1.23 \pm 0.20 $   & $-1.21 \pm 0.06 $  \\
log He/H             & $10.91 \pm 0.01 $ & $10.988 \pm 0.003$  & $10.98 \pm 0.02$   \\
$Y$             &  0.2433  & 0.2760   &  0.2486     \\
$Z$             &  0.0066  & 0.0137   &  0.0122    \\
\enddata

\tablenotetext{a}{\citet{mus99,ven01,pei05}.}
\tablenotetext{b}{\citet{cun94,est04}.}
\tablenotetext{c}{\citet{ags05,chr98}.}
\end{deluxetable}

\subsection{Gaseous mass: $M_{gas}$}  

{From} the H~I measurement by \citet{huc86} and adopting a distance of
495 kpc we obtain an $M({\rm H~I}) = 1.36 \times 10^{8}$ \msun.
{From} the helium abundance presented in Table~\ref{tta} we obtain that
$M({\rm He~I})/M({\rm H~I})$ = 0.33.
The $M({\rm H}_2)$  estimated 
for NGC 6822 by \citet{isr97} is about 10 \% of $M({\rm H~I})$.
Consequently, we will
multiply the $M({\rm H~I}) + M({\rm He~I})$ gaseous mass  by a factor
of 1.1 to take into account the contribution due to  $M({\rm H}_2)$.
{From} the previous considerations we obtain that 
$M_{gas} = (1.98 \pm 0.2) \times 10^{8}$ \msun.

\section{Luminosity Evolution and Star Formation Rate} 

\subsection{Star formation rate: $SFR$ and its normalization of the $M_V$}  

The $SFR$ used in chemical evolution models
is given in general by 
a theoretical parametrization that depends on many physical
properties that vary from object to object 
and in many cases are not well constrained. 
Consequently, in some cases  the adopted  $SFR$ might not be realistic.
Fortunately for NGC 6822 we have a $SFR$ based on observations.

\citet{wyd01,wyd03} based on data obtained with the HST has derived the 
star formation rate ($SFR_i$)
of eight fields in NGC 6822.
Since these 8 fields correspond to regions
of the galaxy with different luminosities and taken together
amount to 20\% of the total luminosity of the object (see bellow), we will 
assume that they are representative of the whole galaxy during its
evolution. Based on the $SFR_i$s 
by Wyder we have constructed a $SFR$ of the whole galaxy.

We have to normalize the results by Wyder to the cosmological  age 
adopted by us. 
Our cosmological model indicates that the galaxy started forming 13.5 Gyr ago,
we have assumed a $SFR = 0$ for the first 1.2
Gyr, therefore the oldest stellar population has a representative age of 12.3 Gyr.

Wyder (2001, 2003) only shows the $SFH$ of region VIII for three assumed galactic ages
 (9, 12 and 15 Gyr),
for the other seven regions he only shows the $SFH$ for an age of the galaxy of 15 Gyr
(note that for the $SFR$ time increases towards the future, while for the $SFH$ time
increases towards the past).
In this paper we have assumed that the differences in the $SFH$ between the ages of
12 Gyr and 15 Gyr are small (as it is the case for region VIII),
and have taken the sum of the $SFH$ of all the regions for an
age of 15 Gyr as representative of the whole galaxy.
In addition we have adopted the $SFH$ in the 15 to 7 Gyr age interval as representative of the
12.3 to 7 Gyr age interval.

The total star formation rate of the eight fields will be called
$SFR(t)_{8f}$ and will be given by

\begin{equation}
SFR(t)_{8f} =  \sum_{i=1}^{8} SFR(t)_i.
\end{equation}

\noindent
With the $SFR(t)_{8f}$ and the GALAXEV program of Bruzual and Charlot (2003)
we obtain an $M_V = -14.28$ at the present time ($t=13.5$ Gyr), which implies that the eight 
HST fields produce 20\% of the luminosity of the galaxy ($M_V = -16.10 \pm 0.1$,
Hodge 1977).
We added to the GALAXEV program an additional subroutine that includes the metallicity
evolution, $Z(t)$ (Bruzual, private communication).
The $Z(t)$ used by us is given by the average of the  [Fe/H] enrichment
histories adopted by Wyder (2003 and private communication)
under the assumption that [Fe/H] is proportional to the  $Z$ value.
This Z(t) function will be called $Z_{Wyder}(t)$.

The $SFR_{total}$ was computed from $SFH$s based on the Wyder code
and was normalized based on the Bruzual-Charlot code.
Both codes are different, but we think that
the procedure used to derive the luminosity evolution is reasonable,
because there are many common ingredients between
GALAXEV and Wyder's codes.
For example: both codes use the same IMF (Salpeter 1955) in a similar mass range
(from 0.1 \msun \ to $100-120$ \msun),
and the same stellar evolutionary isochrones (those of the Padova group
dependent on metallicity).

Assuming that the rest of the galaxy behaves as the HST fields observed by Wyder
we obtain that 

\begin{equation}
SFR_{total}(t) = N_{SFR} \times \sum_{i=1}^{8} SFR(t)_i,
\end{equation}

\noindent
where $N_{SFR}$ is the  normalization constant.
For $N_{SFR}= 5$ we obtain  $M_V = -16.07$ at the present time in excellent agreement
with the observed value ($M_V = -16.10 \pm 0.1$).
In Figure 4 we present the $SFR_{total}$ obtained, this $SFR_{total}$ will be 
assumed in the models.

\begin{figure}[ht]
\plotone{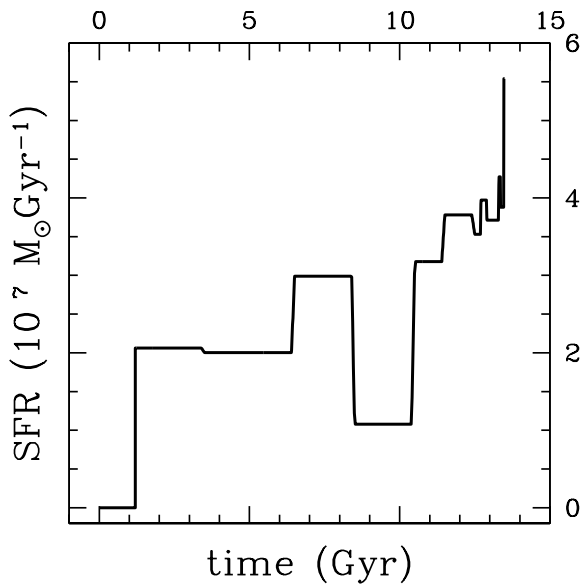}
\caption{
Star Formation Rate of NGC 6822 as a function of time
determined from the star formation histories of the 8 HST fields.
}
\end{figure}

The $M_V$ was obtained with the GALAXEV code under the assumption of Salpeter IMF
and $Z_{Wyder}(t)$. 

To reproduce the chemical evolution of the solar vicinity it has been found that the IMF by 
Kroupa, Tout \& Gilmore (1993, KTG)
is considerable better than the Salpeter IMF (e.g. Carigi et al. 2005).
Therefore in the next section  the chemical evolution models have been computed 
by adopting the KTG IMF. 
To test the robustness of the derived the $SFR$ we have computed 
the $M_V$ value with GALAXEV for the KTG IMF and three different $Z(t)$ functions:
$Z_{Wyder}(t)$, $Z(t)$ of model 7L, and $Z(t)$ of model 8S, our two best chemical evolution models (see section 7).
For $Z_{Wyder}(t)$, $Z(t)$ of model 7L, and $Z(t)$ of model 8S, we obtained 
$M_V = -16.22$, $M_V = -16.18$, and $M_V = -16.14$, respectively;
these values are within one $\sigma$ of the observed $M_V$ value,
and imply that $SFR$ presented in Figure 4 is robust in changes to the IMF and $Z(t)$ functions.

\subsection{Luminosity and color evolution}

In Figure 5 and 6 we present the $M_U$, $M_B$, and $M_V$ absolute magnitudes evolution and
the $J-H$, $J-K$, and $V-J$ colors evolution, respectively,
as a function of time for NGC 6822  given by the GALAXEV code with the KTG IMF and
$Z(t)$ of model 7L.

\begin{figure}[ht]
\plotone{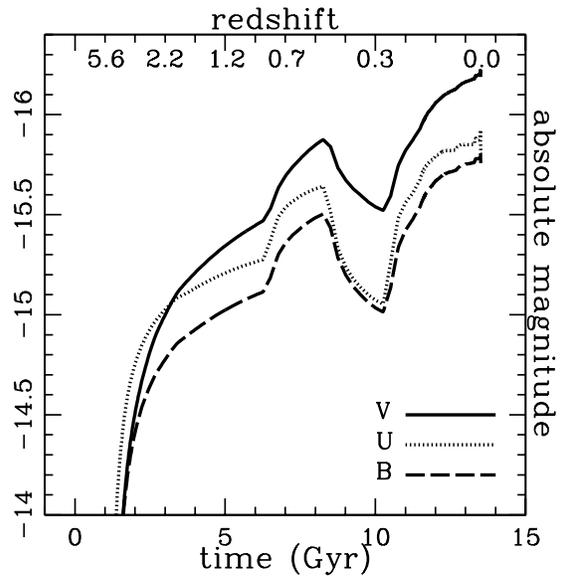}
\caption{
U, B, and V absolute magnitudes for NGC 6822 as a function of time.
$M_U$, $M_B$, and $M_V$ were computed from the $SFR$ shown in Fig. 4 using
the GALAXEV code with metallicity evolution of model 7L (our best chemical
evolution model) and the  KTG IMF.
}
\end{figure}

In Table 2 we compare the observed magnitudes and colors 
with those predicted by different models for the present time (13.5 Gyr).
The magnitudes $M_B$ and $M_V$ are taken from 
Wyder (2001), while the colors
 $M_J$ - $M_H$ and $M_J$ - $M_K$ from Jarret et al. (2003).
The $U-B$ color is from Hodge (1977)
and the reddening correction for $U-B$ was taken from \citet{sch98}.
The agreement between the model predictions and observations is excellent
 for the $M_U$, $M_B$, and $ M_V$ magnitudes
and the $M_J$ - $M_H$ and $M_J$ - $M_K$ color indexes
(see also Figure 6).

\begin{figure}[ht]
\plotone{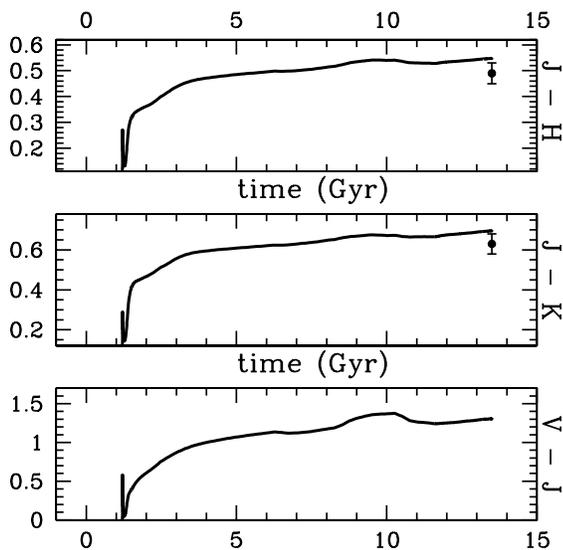}
\caption{
Same as Fig 5, but for the $J-H$, $J-K$, and $V-J$ colors.
The observational constraints $M_J$ - $M_H$ and $M_J$ - $M_K$
 are also presented for comparison (Jarret et al. 2003).}
\end{figure}

It is not possible to compare the visual magnitudes with
the infrared magnitudes because the infrared observations did not include the whole galaxy
(Jarrett et al. 2003).
By adopting the $M_V$ - $M_J$ color of the $Z(t)$ 7L model, we can determine the fraction of the integrated
 light of the galaxy observed in the infrared. 
The photometric evolution model predicts $M_V$ - $M_J= 1.32$ and
$M_J= -17.50$.
On the other hand the observed $M_J$ value by Jarrett et al. (2003) amounts to $-16.2$
a value 1.3 mag fainter than predicted by the model.
This difference implies that the aperture used by Jarrett et al. (2003) included only 
30 \% of the integrated light of the galaxy, 
in agreement with the comment in their section 6.5   
where they mention that in general they observed only from 20 \% to 50 \% of
the integrated light of dwarf irregular galaxies.

{From} Table 2 it can be seen that the different models agree with the observations
at one $\sigma$ level. In other words the absolute luminosity in different bands 
as well as the color indexes are
almost independent of the three different $Z(t)$ functions as well as the two IMFs
used and consequently that the $SFR$ presented in Fig. 4
is robust and can be used for all the chemical evolution models.
That is the $SFR$ has been derived from observations
and it will be kept fixed for all the chemical evolution models.

\begin{deluxetable}{cccccc}
\tablecaption{Present-day Photometric Properties for different Initial Mass Fuctions and Metallicity Evolutions
\label{phot}}
\tablewidth{0pt}
\tablehead{
\colhead{IMF} &
\colhead{Salpeter} &
\colhead{KTG} &
\colhead{KTG} &
\colhead{KTG} &
\colhead{Observations}\\
\colhead{$Z(t)$} &
\colhead{Wyder} &
\colhead{Wyder} &
\colhead{Model 7L} &
\colhead{Model 8S} &
}
\startdata
$ M_U$ & -15.97 & -15.91 & -15.82 & -15.74 & -15.83 $\pm$ 0.16 \\
$ M_B$ & -15.74 & -15.81 & -15.76 & -15.70 & -15.60 $\pm$ 0.13 \\
$ M_V$ & -16.07 & -16.22 & -16.18 & -16.14 & -16.10 $\pm$ 0.10 \\
& & \\
$ M_J$ - $ M_H$ &  0.52 & 0.53 & 0.55  & 0.56 & 0.49 $\pm$ 0.04 \\
$ M_J$ - $ M_K$ &  0.66 & 0.66 & 0.70  & 0.72 & 0.63 $\pm$ 0.05 \\
$ M_V$ - $ M_J$ &  1.18 & 1.26 & 1.32  & 1.35 & --- \\
& & \\
$Z_{NOW}$ & 0.0024 & 0.0024 & 0.0069 & 0.0084 & 0.0066 $\pm$ 0.0011\\ 
\enddata
\end{deluxetable}

\section{Galactic Winds}

A common hypothesis in chemical evolution models of dwarf galaxies is
to assume that these galaxies expelled part of their gas during their lifetime:
a galactic wind is mostly introduced in the modelling to explain the low chemical abundances 
found in these type of galaxies and NGC~6822 seems to be no exception. From
an observational point of view the amount of evidence of outflows in star forming galaxies 
is increasing with time and we refer the reader to \citet{vcbh05} for an updated 
discussion of galactic winds. On the other hand, from a theoretical point of view, 
numerical simulations show that a blow-away outflow situation is not rather 
uncommon phenomenon in low-mass galaxies \citep[e.g][]{fuj04}. 

In order to see if NGC 6822 could have had a
galactic wind during its lifetime, specially at high redshift, we have computed
its present date binding energy as well as the thermal energy history of the gas.
In Figure 7 we present the thermal energy of the gas as a function of time assuming no outflows.

The thermal energy was computed as follows:
we define $E_{THER}$ as the thermal energy of the gas from supernovae,
\begin{equation}
E_{THER}(t) = \int_0^t{\epsilon(t-T)RSN(T)dT}
\end{equation}
where $RSN$ is the supernovae rate and includes types Ia and II SNe.
The $\epsilon(t)$ function represents the evolution of the thermal energy in the hot, dilute interior
of the supernovae remnants and is given by

$$\epsilon(t)=\cases{0.72\epsilon_0 & if $t \le t_C$, \cr
                            0.22 \epsilon_0 (t/t_C)^{-0.62} &  if $t > t_C$, \cr} $$
where $\epsilon_0=10^{51}$ erg and $t_C$ is a cooling time scale taken from Cox (1972).
For further details see Carigi, Hern\'andez \& Gilmore (2002).

Also in Fig. 7  we show the binding energy of NGC 6822 for the present gas content
based on the following equation:

\begin{equation}
E_{BIND} = G \frac{M_{gas} M_{total}}{R_{vir}}.
\end{equation}

As can be seen from Figure 7 the thermal energy available, released during the lifetime of the galaxy,
is higher than the present binding energy of the gaseous content of the galaxy, which implies that
part of the thermal energy was lost by the galaxy in the past. This result is in agreement with
the chemical evolution models that require a substantial gas outflow to match the present gaseous content
with the observed O/H value (see section 7).

As we mention above, the computation of the thermal energy content of 
NGC 6822 is based on the cooling scheme by Cox (1972). A more
elaborated treatment of the SN remnant cooling has been presented by 
Tantalo et al. (1998). They computed the cooling time as a function of 
gas metallicity and found that for metallicities below solar it is larger 
than that predicted by Cox. Thus, in absence of winds, the thermal energy 
in NGC 6822 should lie above the one shown in Fig. 7.

\begin{figure}[ht]
\plotone{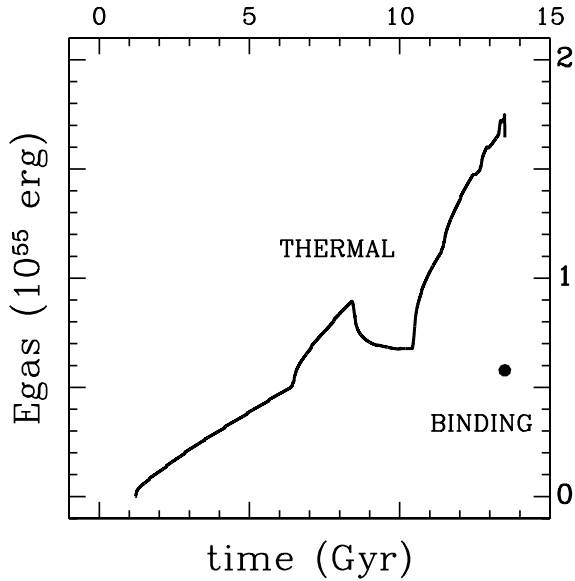}
\caption{
Gaseous thermal energy predicted by $SFR$ under the assumption of no gaseous outflow from the galaxy based on
the formulation by Cox (1972). We also show the binding energy of the present gas content of the galaxy.
The difference between the predicted thermal energy and the binding energy at present implies that
the galaxy must have lost thermal energy due to outflows in the past. 
}
\end{figure}

The outflows needed to explain the excess thermal energy predicted by the observed $SFR$
and by the chemical evolution model presented in section 6 very likely occurred in the first
few Gyr of evolution ($t< 5$ Gyr) because:
i) To propitiate an efficient blow away process strong bursts of star formation are needed \citep{ten03},
{from} the $SFR$ it is not possible to rule out the presence of several outburst processes
at early times ($t< 5$ Gyr), because the $SFR$ has been averaged at long time intervals ($\sim$ 2.5 Gyr),
and it is possible for one or two large bursts of star formation,
to go unnoticed in the derived $SFR$,
on the other hand, at later times the $SFR$ has been averaged for shorter time intervals ($\sim$ 0.5 Gyr),
and it is more difficult for a high $SFR$ to go unnoticed;
ii) in the past, in a hierarchical cosmology, 
the potential wells of galaxies, for a given present-day mass, 
are shallower; 
iii) strong star formation events are 
expected  to occur because of the higher galaxy
merger rate (e.g., Maller et al. 2005); 
iv) after an outflow occurs, $M_{gas}$ gets diminished and the possibility
of a second outflow due to a burst of star formation
increases because there is a lower amount of gas in the ISM to reduce the
velocity of the SN ejecta, and
v) the low metallicity at early times increases the cooling time relative to the classical Cox (1972)
value, and therefore increases the thermal energy available propitiating the advent of galactic winds.
As can be seen from our chemical evolution models, for example models 7L and 8S (see section 7)
the $Z$ value for $t < 5$ Gyr is about 0.001, thirteen times smaller than solar. 

It is beyond the scope of this paper to study the outflow rate as a function of time.

\section{Chemical Evolution Models}  

We have computed chemical evolution models with a very low number of free 
parameters. The cosmological context provides the accretion rate, and
the luminosity model provides the $SFR$ as a function of time, 
both in an independent way.
The remaining ingredients related with stellar populations: the IMF,
the stellar yields, and the percentage of binary systems
which explode as SNIa, are inferred from a successful
chemical evolution model of the solar vicinity and
the Galactic disk by \citet{car05}.

To construct the models for NGC 6822, we have relied upon
the model of the solar vicinity 
by \citet{car05}, consequently we have adopted the following assumptions:
i) the IMF proposed by Kroupa, Tout, \& Gilmore (1993, KTG) for the 0.01 - 80  \msun \ range,
ii) for massive stars, those with $ m > 8$ \msun, we have used 
yields that depend on the initial metallicity as follows,
for $Z_i=0$ yields by Chieffi \& Limongi (2002), 
for $Z_i= 1 \times 10^{-5}$ and  0.004 yields by \citet{mm02}, 
and for $Z_i= 0.02$ yields by Maeder (1992),
iii) for low-and-intermediate-mass stars, those with $ m < 8$ \msun \
and from $Z_i=0.004$ to $Z_i=0.02$   we have used the yields by 
Marigo, Bressan, \& Chiosi (1996, 1998), and \citet{pcb98},
iv) for Type Ia supernovae, we have used the yields for $Z_i= 0.02$ 
by Thielemann et al. (1993), these yields are almost independent of metallicity,
and
v) the fraction of binary systems with total mass
between $ 3 < m(\msun) < 16 $ that becomes SNIa ($Abin$) amounts to 0.05.
For each set of yields, linear interpolations for different stellar masses
and metallicities were made.
For metallicities higher or lower than those available we adopted the yields predicted by
the highest or lowest $Z$ available, respectively.

In some models, we will adopt
two different types of outflows:
i) we call ``selective outflows" those constituted only by
material ejected by supernovae of Type II and Type Ia,
in this case all the material  ejected by massive stars (Type II SNe)
is  lost to the IGM, and
ii) we call ``well mixed outflows" those that reach the IGM after the ejecta
by MS, LIMS, and Type Ia SNe have been completely mixed with the ambient ISM.
We have assumed that the rate  of the well mixed outflow
is proportional to the $SFR$ and that the proportionality  coefficient is
constant during the presence of the outflow.
With both types of outflows the galaxy loses material that never
gets back. The duration of the outflow is  another free parameter.

We computed a large set of chemical evolution models.
{From} these, we will discuss in detail 15 models
in sections 7.1 and 7.2.
All models  evolve with the mean MAH
computed in section 3.3. Seven of the 15
models follow a baryonic mass accretion history
that does not assume large-scale heating of the gas,
while the other eight models do.
In some models we included outflows of baryonic matter.
All models  fit the observed colors evolution  in Section 5.2,
and twelve of them have the same values for $M_{sub}$,  $M_{stars}$, and $M_{rem}$,
at the present time, these values  amount  to $3.48 \times 10^7$ \msun, $1.73 \times 10^8$ \msun,
and $2.31 \times 10^7$ \msun.
The  exceptions are Model 7L, 7S, and 8S, that
have different upper mass limits for the IMF, for these models the values of
$M_{sub}$,  $M_{stars}$, and $M_{rem}$
are practically the same values than those predicted by the other models and
they will be presented in sections 6.1 and 6.2.
The  $M_{bar}$  determined from the observed gaseous mass
and the other terms present in eq. 2
amounts to $4.3 \times 10^8$ \msun.

\subsection{Large infall models}

We have computed several models that adopt all the ingredients presented in the previous sections,
specially in Sections 3, 5.1 and 7.
In particular, in Models 1L -- 7L we consider that the baryonic mass accreted by NGC 6822
is that suggested by the cosmological context without large-scale shock heating.

In Table 3 we present the main characteristics of seven models.
Columns 2 and 3 describe the type of outflow and its duration.
Columns 4 to 8  present five properties predicted by the models:
$M_{gas}$, O/H, C/O, N/O, and Fe/O, as well as their observed values.
The $M_{gas}$ and the O/H values are the most critical observational
constraints for NGC 6822, therefore
a model that fits the C/O, N/O and Fe/O observed values,
but that does not fit the observed O/H value, is a poor model because it does not fit
the C/H, N/H, and Fe/H observed values.

In Model 1L we assume that all the baryons available after the reionization fall into
the galaxy and remain in the galaxy during its whole evolution.
In Figure 8 we show the evolution history of six properties of the model up to now.
Model 1L grossly fails to fit $M_{gas}$ and O/H, since it predicts 4.7 times more gas than observed
and a lower O/H  value than observed by 5 $\sigma$
(see Table 3 and Figure 8).  

\begin{figure}[ht]
\plotone{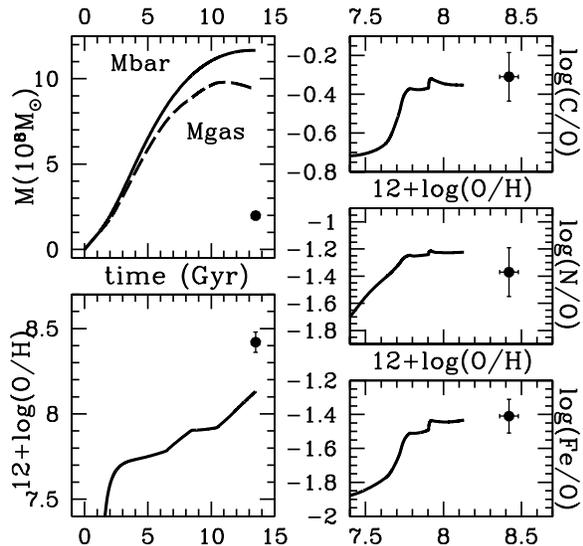}
\caption{
Model 1L.
Mass formation history and predicted chemical evolution.
This model assumes that all the gas available after the reionization falls into the galaxy
and does not include outflows.
The present-day  observational constraints (filled circles) are also presented for comparison.
The model fails drastically.
}
\end{figure}

In order to reduce the enormous difference between the predicted $M_{gas}$
value by Model 1L and the observed value we have computed
models with well mixed outflows during the early history of the galaxy.

Model 2L is built to reproduce the $M_{gas}$ observed value under the assumption
of an instantaneous well mixed outflow.
This event had to occur at 6.3 Gyr, because at that time
the $M_{gas}$ of the galaxy amounts to the difference between
the observed and the predicted $M_{gas}$ by Model 1L (7.38 $\times 10^8$ \msun).
This model reproduces $M_{gas}$, C/O and Fe/O, but the
O/H and N/O values are higher than observed
by 2.0 $\sigma$ and 1.2 $\sigma$ respectively (see Table 3).
O/H is higher because according to the $SFR$  most of the stars are
formed after the outflow event and the O ejected by the stars to the ISM
is kept by the galaxy

Based on the models by \citet{fuj04} we find  that in order to produce
an instantaneous well mixed wind in NGC~6822 at 6.3 Gyr, when its total mass is
$1.7 \times 10^{10} \msun$, about 10\% of the gas should be converted into stars.
This amount of gas corresponds to $\sim 6 \times 10^7$ \msun \
which according to the derived $SFR$ is 2.5 times higher than the mass converted
into stars by NGC 6822 in an interval of  1 Gyr centered at 5.8 Gyr (see Figure 4).
Consequently, we consider Model 2L to be unlikely.

Model 3L is built to reproduce the O/H observed value under the assumption of a well
mixed outflow that starts at the same time than the star formation ($t= 1.2$ Gyr).
To reproduce the observed O/H value the model needs the outflow phase to last 4.0 Gyr.
During the outflow phase the mass lost to the IGM is
the maximum possible, that means that the gas left is the minimum required to maintain
the $SFR$ given in Fig. 4.

Model 3L reproduces all the observational constraints related with chemical
abundances, but not the $M_{gas}$ observed value:
the $M_{gas}$ predicted is higher than observed,
by a factor of 1.65 (8 $\sigma$) (see Table 3 and Figure 9).
The gas mass lost is 6.09 $\times 10^8$ \msun,                                         
8.5 times the mass of stars formed during the first 4.0 Gyr.
During  the wind phase the outflow rate is equal to 7.51 times the $SFR$.
If it turns out that 
the observed gas mass value has been underestimated (which is unlikely 
since most of the gas is in the form of HI that is observed), 
this model might become relevant.

\begin{figure}[ht]
\plotone{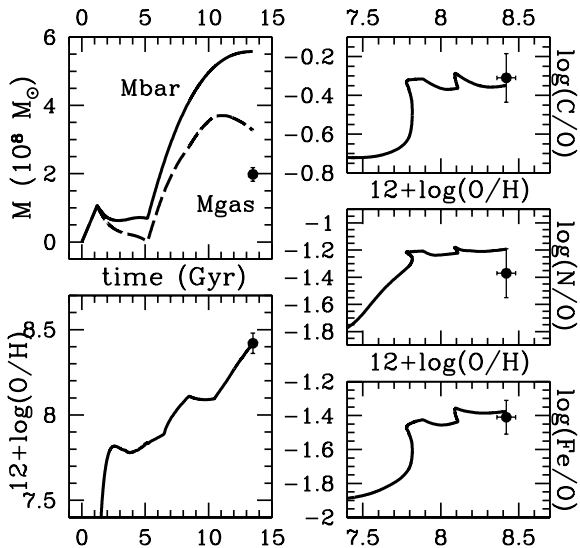}
\caption{
Model 3L.
Mass formation history and predicted chemical evolution. This model
assumes the gas mass assembly history of the large infall models
and  includes a well mixed outflow during the first 4 Gyr of the star formation.
}
\end{figure}

Model 4L is built to reproduce the $M_{gas}$ observed value under the assumption
of a well mixed outflow.
This model requires to lose more gas than Model 3L, therefore the duration of
the outflow has to be longer.
To fit $M_{gas}$ we need the outflow to last 5.1 Gyr.
During the outflow phase the mass lost to the IGM is
the maximum possible, that means that the gas left is the minimum required to maintain
the $SFR$. During the wind phase the outflow rate is equal to 7.11 times the $SFR$.
The gas mass lost is 7.35 $\times 10^8$ \msun,
8.0 times the mass of stars formed  during the first 5.1 Gyr of the star formation.
The predictions of this model are identical to those of Model 2L (see Table 3),
from $\sim$ 7 Gyr up to now, because:
i) both models at $\sim$ 5 Gyr had lost practically
all of the gas and  started accumulating gas again at the same rate, and
ii) both models had  the
the same $SFR$  after 5 Gyr.
In other words, after the end of the outflow phase both models have
the same stellar populations that pollute equally the same amount of gas.

Model 4L agrees with the observed $M_{gas}$ value
but overproduces  O, therefore we will modify
 Model 4L to fit the O/H ratio and the other observational constraints
by adopting  extra assumptions in Models 5L, 6L, and 7L.

In Model 5L we have assumed that 0.5 Gyr ago a primordial gaseous cloud of
0.34 $\times 10^8$ \msun \ was accreted by  NGC 6822.
We have assumed the least massive cloud needed 
to reproduce the  O/H value.
With this extra infall, the model can reproduce all the observational
constraints related to chemical abundances with the  exception of N/O, 
but the predicted $M_{gas}$
is higher by a factor of 1.2 (see Table 3).
The accretion of a huge gaseous cloud of primordial material
recently is more unlikely than at  earlier times.
If we assume the accretion of a primordial cloud at earlier times
($t<13$  Gyr), 
the model would require a more massive cloud, increasing the 
disagreement with the observed $M_{gas}$.

To assume the infall of a cloud of $3.4\times10^7$ \msun \
of primordial gas at $t=13$ Gyr is unlikely for the following reasons:
i) this gaseous cloud has about 8 \% of the baryonic mass of the galaxy
and therefore could have formed part of the NGC 6822 satellite system,
ii) at such late time a gas cloud of this mass could have had time to form stars and
these stars could have changed the primordial chemical composition 
of the gas cloud (according to De Block \& Walter (2005)
the disk of NGC~6822 does have an embedded gaseous
cloud with comparable mass but with an old stellar population), and
iii) the total gaseous mass accreted by the galaxy in the last 0.5 Gyr
according to the MAH amounts only to $5\times10^5$ \msun. 
Consequently, we consider Model 5L unlikely.

In Model 6L we have assumed a second outflow of the selective type,
during the last 1 Gyr, when the $SFR$ is most important.
In this outflow, all the  material ejected by
SNII and SNIa is lost to the IGM.
If the SNIa ejecta remain in the galaxy,
the predicted Fe/O value becomes $-1.30$ dex, in marginally
 agreement with observations (see Table 3).
Model 6L is unlikely because between 12.5 and 13.5 Gyr
the $SFR$ does not indicate the presence of a major starburst
(see Figure 4) needed to drive the outflow,
and the potential gravitational well is largest.
Moreover, we think that it is unlikely that the 
material ejected by every SNe in the last Gyr escapes to the IGM.

In Model 7L
we have reduced the IMF mass upper limit, $m_{up}$,  from 80 \msun \ to 60 \msun.
When $m_{up}$ is reduced, the percentage of MS, main producers of O,
is decreased by a factor of 1.18, and the predicted O/H value is reduced by 0.12 dex.
This fact produces an increase of the  N/O and Fe/O predicted values to
-1.06 dex and -1.24 dex, respectively, more than 1 $\sigma$ away from the observed values.
The Fe/O value can be adjusted by assuming a lower
fraction of binary systems that become SNIa ($Abin$).
For this reason we have reduced $Abin$ from 0.05
(the value assumed in the  rest of the L models) to 0.02.
Model 7L fits all the observational constraints with the exception of the N/O ratio
(see Figure 10 and Table 3). 

\begin{figure}[ht]
\plotone{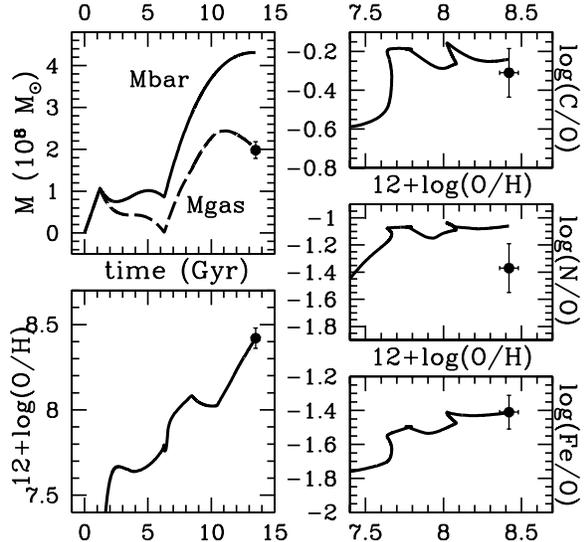}
\caption{
Model 7L.
Mass formation history and predicted chemical evolution.
This model assumes the gas mass assembly history of the large infall models
and includes a well mixed outflow during the initial
5.1 Gyr of the star formation.
The IMF $m_{up}=60$ \msun,  and the fraction of binary systems that becomes SNIa
is 2 \%.
}
\end{figure}

\begin{deluxetable}{ccccccccc}
\rotate
\tablecaption{Inputs and outputs of large infall models 
\label{model1}}
\tablewidth{0pt}
\tablehead{
\colhead{Model} &
\multicolumn{2}{c}{Wind} &
\colhead{$M_{gas}$} &
\multicolumn{4}{c}{Abundances} \\
&
\colhead{Type} &
\colhead{Duration} &
\colhead{($108$ \msun)} &
\colhead{12+log(O/H)} &
\colhead{log(C/O)} &
\colhead{log(N/O)} &
\colhead{log(Fe/O)} \\
 & &
\colhead{(Gyr)} &
& & & &
}
\startdata
 1L &         --- &  --- & 9.36 & 8.13 & -0.35 & -1.22 & -1.43 \\
 2L &  Well Mixed &  0.1 & 1.99 & 8.54 & -0.33 & -1.16 & -1.35 \\
 3L &  Well Mixed & 4.0  & 3.27 & 8.42 & -0.35 & -1.19 & -1.38 \\
 4L &  Well Mixed &  5.1 & 2.01 & 8.54 & -0.33 & -1.16 & -1.35 \\
 5L \tablenotemark{a}&  Well Mixed &  5.1 & 2.34 & 8.48  & -0.34 & -1.17 & -1.35 \\
 6L \tablenotemark{b}&  Well Mixed &  5.1 & 1.98 & 8.42 & -0.27 & -1.09 & -1.37 \\
 7L \tablenotemark{c}&  Well Mixed &  5.1 & 2.01 & 8.42 & -0.24 & -1.06 & -1.41 \\
&  & & & & & & \\
Obs. &  & &  1.98 $\pm$ 0.20 & 8.42 $\pm$ 0.06 & -0.31 $\pm$ 0.12 & -1.37 $\pm$ 0.18 & -1.41 $
\pm$ 0.10 \\
\enddata
\tablenotetext{a}{Extra $3.4 \times 107$ \msun \ of primordial gas accreted at 13.0 Gyr. }
\tablenotetext{b}{All the  material expelled by Type II and Ia SNe during the last Gyr is ejected
to the IGM. }
\tablenotetext{c}{The upper mass limit for the IMF amounts to 60 \msun.
The fraction of binary systems that become SNIa amounts to 0.02.
}
\end{deluxetable}

With $m_{up}=60$ \msun, and $Abin = 0.02$, the stellar populations are not the same than
those in the solar vicinity. With these values the model by Akerman et al. (2005) or Carigi et al.
(2005)  cannot reproduce the chemical observational
constraints for the solar vicinity,
implying that the IMFs of the solar vicinity and of NGC 6822 are not the same.
Other authors have also argued in favor of smaller $m_{up}$ values for dwarf galaxies,
for example,
Weidner \& Kroupa (2005) suggest that $m_{up}$ is lower
 in dwarf galaxies than in the solar vicinity;
the smaller $m_{up}$ implies that the number of Type II SN per generation
of stars has to be smaller than in the solar vicinity.
Their results depend on the $SFR$. For a continuous $SFR$ during the last 14 Gyr,
their IMF slope is steeper than the Salpeter IMF for massive stars,
as in the case of the IMF assumed in this paper,
and has an $m_{up} \sim 10$ \msun. 
For a single 100 Myr burst their slope is like the Salpeter's one
and has an $m_{up} \sim 100$ \msun.
The $SFR$ computed by us is nearly constant during the first 5 Gyr and
then shows several bursts. Based on the work by Weidner \& Kroupa
the IMF adopted by us should be dependent on time, but that is out of the
scope of this paper.
We consider that a constant IMF with $m_{up}=60$ \msun \ is a good approximation to
the IMF proposed by Weidner \& Kroupa (2005).
Other authors also argue that galaxies with a lower $SFR$ have a lower $m_{up}$ value
in agreement with our suggestion
\citep[e.g.][and references therein]{gp05}.

The change in $m_{up}$ from 80 \msun \ to 60 \msun \ affects the O/H evolution, but it does
not affect the photometric evolution nor the thermal evolution.
We have computed a photometric evolution model with $m_{up}=60$ \msun \ and obtained
$M_U=-15.82$, $M_B=-15.77$,  $M_V=-16.19$, $M_J-M_H=0.55$, $M_J-M_K=0.70$, and $M_V-M_J=1.32$
 practically identical to those
derived with $m_{up}=80$ \msun \ presented in Table 2, 
therefore the change in the $SFR$ due to
the reduction of $m_{up}$ from 80 \msun \ to 60 \msun \ is negligible.
Similarly, the gaseous thermal energy produced by the $SFR$ under the assumption of no outflows diminishes
from 1.65 $\times 10^{55}$ erg to 1.63 $\times 10^{55}$ erg when $m_{up}$ changes from 80 to 60 \msun.
 
For Model 7L the current values of 
 $M_{sub}$,  $M_{stars}$, and $M_{rem}$,
amount to $3.49 \times 10^7$ \msun, $1.73 \times 10^8$ \msun,
and $2.31 \times 10^7$ \msun, respectively.

It can be shown that the use of the Salpeter IMF 
would imply a smaller $m_{up}$ than that required by KTG IMF 
to fit the observed O/H value.
Consequently, since all of the O is produced by massive stars and
only a fraction of the C is produced by them,  the C/O value would become 
considerably larger than observed (e.g. see model 7S).

In Table 4 we present two carbon  budgets:
the stellar production one, and the interstellar medium one
for L models. The production is due to
massive stars (MS), low and intermediate mass stars (LIMS), and SNIa. Also
in Table 4 we present for comparison the carbon budgets for the
solar vicinity (Carigi et al. 2005). 
For Model 1L both budgets agree because there is no outflow present, a similar
situation prevails for the solar vicinity.
For Models 4L, 6L,  and 7L the contribution of massive stars to the
C abundances in the ISM is smaller than the stellar production
due to outflows to the IGM. 
Since for Models 4L and 6L the IMF and
the $SFR$ are the same, the higher contribution of MS to the C
budget is due to the dependence of the C yields in the O/H ratio,
which is higher for Model 4L than for Model 6L. For MS the C
yields increase with metallicity due to stellar winds, while for LIMS the
C yields decrease with metallicity (Carigi et al. 2005). Since for Models 6L and 7L
the $SFR$ and the O/H values are the same, the higher contribution
of MS to the C budget is due to the difference in the IMFs, while for
Model 6L $m_{up}$ amounts to 80 \msun \ for Model 7L it only amounts to 60 \msun.
Finally the larger contribution of MS to the C budget in the solar
vicinity than in Model 1L is due to the evolution of the O/H values,
and as mentioned before the higher O/H values produce higher C yields
for MS and lower ones for LIMS.

The difference among the carbon budgets of the computed models for
NGC 6822 is small and implies that the result obtained, in the sense that
the LIMS produce about 63 \% of the C abundance and that the MS
produce about 36 \% of the C abundance, is a robust one. 

\begin{deluxetable}{cccc}
\tablecaption{Carbon budget for large infall models\tablenotemark{a}
\label{cprodr}}
\tablewidth{0pt}
\tablehead{
&
\multicolumn{3}{c}{Contribution (per cent)} \\
\colhead{Model } &
\colhead{MS} &
\colhead{LIMS} &
\colhead{SNIa}
}
\startdata
1L & 36.3 & 61.4 & 2.3 \\
4L & 38.2 & 59.5 & 2.3 \\
6L & 37.7 & 60.1 & 2.3 \\
7L & 35.9 & 63.2 & 0.9 \\
Solar Vicinity & 48.2 & 49.8 & 2.0 \\
& & & \\
\multicolumn{4}{c}{Remaining values in the ISM} \\
1L & 36.3 & 61.4 & 2.3 \\
4L & 37.2 & 60.4 & 2.4 \\
6L & 30.8 & 67.1 & 2.1 \\
7L & 34.2 & 64.8 & 1.0 \\
Solar Vicinity & 48.2 & 49.8 & 2.0 \\
\enddata
\tablenotetext{a}{
Percentage of C in the ISM
produced by different types of stars
over a period of 13 Gyr.
}
\end{deluxetable}

\subsection{Small infall models}

For the S models 
we consider the baryonic mass aggregation history
obtained in section 3.4 under the assumption of 
large-scale gas shock heating, that prevents
a fraction of the gas to be accreted by the galaxy.
Specifically, only 60 \% of the baryons available
after taking into account the effect
of reionization (see Figures 2 and 3, dotted lines)
fall into the galaxy. The remaining 40 \% of the baryons
are heated by the collapse of large-scale structures  (filaments and
pancakes) and are never accreted by the galaxy.

In Table 5 we present the main characteristics of eight S models.
Columns 2 and 3 describe the type of outflow and its duration.
Columns 4 to 8  present five properties predicted by the models.

In Model 1S we assume that the gas that falls
into the galaxy remains in the galaxy during its whole evolution.
In Figure 11 we show the evolution history of six properties of the model up to the
present time.
Model 1S fails to fit $M_{gas}$, since it predicts 2.4 times more gas than observed,
but reproduces all the observational constraints related with chemical
abundances (see Table 5 and Figure 11).
If the observed $M_{gas}$ value  were higher than the adopted in this paper,
this model would be one of the best and would be the only successful model
without outflow.

\begin{figure}[ht]
\plotone{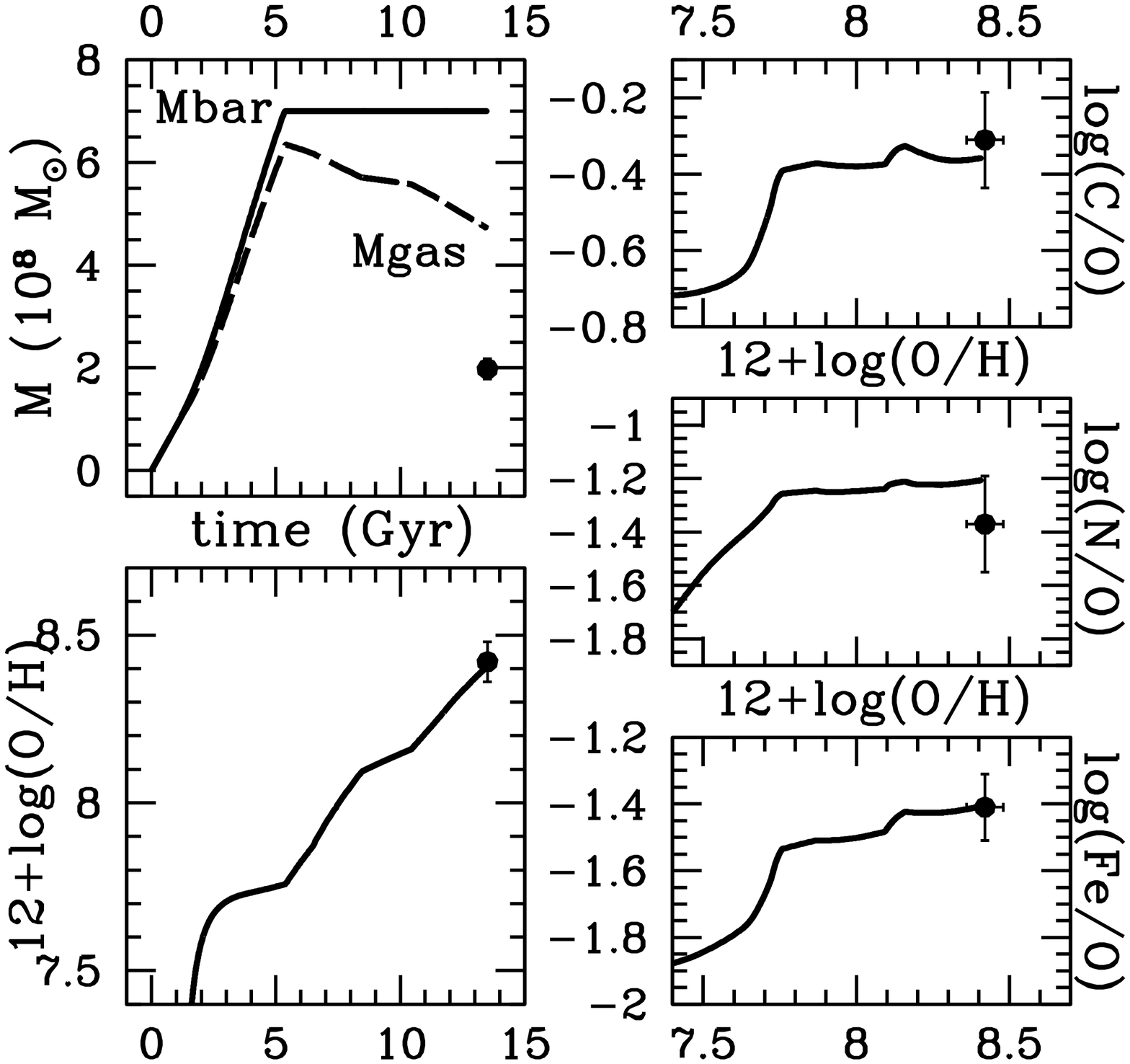}
\caption{
Model 1S.
Mass formation history and predicted chemical evolution.
This model assumes that 40 \% of the gas available after the reonization does not fall 
into the galaxy
and this model does not include outflows.
The present-day  observational constraints are also presented for comparison.
The model fails to reproduce $M_{gas}$.
}
\end{figure}

In order to reduce the difference between the predicted  and observed $M_{gas}$
presented by Model 1S we have computed models that include well-mixed outflows
during the early history of the galaxy.

Model 2S is built to reproduce the observed $M_{gas}$ under the assumption
of an instantaneous well mixed outflow.
This event had to occur at 3.1 Gyr, because at that time
the $M_{gas}$ of the galaxy amounts to the difference between
the observed and the predicted $M_{gas}$ by Model 1S (2.73 $\times 10^8$ \msun).
This model reproduces $M_{gas}$ but the
O/H value is higher than observed
by 3.7 $\sigma$ (see Table 5). The C/H, N/H and Fe/H ratios are not reproduced.

Model 3S is built to reproduce the O/H observed value under the assumption of a well
mixed outflow that starts at the same time than the star formation.
To reproduce the observed O/H value  the outflow phase has to last only 0.1 Gyr.
This model is very similar to Model 1S.
During the wind phase the outflow rate is equal to 10.0 times the $SFR$.
This model reproduces all the observational constraints related with chemical
abundances, but not the $M_{gas}$ observed value (see Table 5).

Model 4S is built to reproduce  $M_{gas}$ value under the assumption
of an early well mixed outflow that lasts 1.6 Gyr.
Since Model 4S accretes less gas than Model 4L, it requires an outflow
during a shorter amount of time to reproduce the $M_{gas}$ observed.
During the outflow phase the mass lost to the IGM is
the maximum possible, that means that the gas left is the minimum required to maintain
the $SFR$.

During the wind phase the outflow rate of Model 4S is equal to 8.56 times the $SFR$.
The gas mass lost in the outflow amounts to 2.73 $\times 10^8$ \msun,
9.5 times the mass of stars formed  during the first 1.6 Gyr of the star formation.
The predictions of this model are identical to those of Model 2S (see Table 5),
from $\sim$ 3 Gyr up to now, because both models at $\sim$ 3 Gyr had lost practically
all of their gas and  started accumulating gas again at the same rate and with the
the same $SFR$ during their later evolution.
In other words, after the end of the outflow phase both models have
the same stellar populations that pollute equally the same amount of gas.

In Model 4S  accretion stops  after 5 Gyr, when the $SFR$ is important.
Therefore, its predicted O/H value is higher than that obtained by Model 4L
because the chemical elements produced by the stars after t= 2.8 Gyr are ejected into an ISM
which is no longer diluted by  primordial infalling gas.
To lower the O/H value of Model 4S we have produced Models 5S -- 7S 
that require even more extreme extra assumptions
than those needed for Models 5L -- 7L.

In Model 5S 
in order to reproduce the O/H value with the least massive cloud
we have assumed that  0.5 Gyr ago a primordial gaseous cloud of
1.2 $\times 10^8$ \msun \ was captured by NGC 6822.
With this extra infall, the model can reproduce all the observational
constraints related to chemical abundances with the exception of N/O,
but the predicted $M_{gas}$ is higher by a factor of 1.6 (see Table 5).
The accretion of a huge gaseous cloud of primordial material
recently is more unlikely than at  earlier times.
If we assume an accretion of a primordial cloud at $t < 13 $ Gyr,
with the same $SFR$ (important at recent times),
the model will require  a more massive cloud producing a larger
disagreement with the observed $M_{gas}$. We consider Model 5S
more unlikely than Model 5L because the mass of the infalling cloud
has to be larger and the disagreement with the observed $M_{gas}$
becomes also larger. 

For Model 6S we have assumed a second outflow
of the selective type during the last 2.5 Gyr, when the $SFR$ is most important.
In this outflow, 100 \% of the  material ejected by
all SNII and SNIa is lost to the IGM.
This model predicts higher C/O and N/O values (see Table 5) because the galaxy keeps
the C and N produced by LIMS.
Model 6S is unlikely because between 11.0 and 13.5 Gyr
the $SFR$ does not indicate the presence of a major starburst
(see Figure 4) needed to drive the outflow,
and the potential gravitational well is largest.
Moreover, Model 6S is more unlikely than Model 6L because it requires a recent rich outflow of longer duration.

\begin{deluxetable}{ccccccccc}
\rotate
\tablecaption{Inputs and outputs of small infall models
\label{model2}}
\tablewidth{0pt}
\tablehead{
\colhead{Model} &
\multicolumn{2}{c}{Wind} &
\colhead{$M_{gas}$} &
\multicolumn{4}{c}{Abundances} \\
&
\colhead{Type} &
\colhead{Duration} &
\colhead{($108$ \msun)} &
\colhead{12+log(O/H)} &
\colhead{log(C/O)} &
\colhead{log(N/O)} &
\colhead{log(Fe/O)} \\
 & &
\colhead{(Gyr)} &
& & & &
}
\startdata
 1S &         --- &  --- & 4.71 & 8.41 & -0.36 & -1.21 & -1.41 \\
 2S &  Well Mixed &  0.1 & 1.99 & 8.64 & -0.29 & -1.13 & -1.35 \\
 3S &  Well Mixed &  0.1 & 4.51 & 8.42 & -0.36 & -1.20 & -1.41 \\
 4S &  Well Mixed &  1.6 & 1.98 & 8.64 & -0.29 & -1.13 & -1.35 \\
 5S \tablenotemark{a}&  Well Mixed &  1.6 & 3.18 & 8.48 & -0.30 & -1.14 & -1.36 \\
 6S \tablenotemark{b}&  Well Mixed &  1.6 & 1.92 & 8.42 & -0.18 & -0.99 & -1.38 \\
 7S \tablenotemark{c}&  Well Mixed &  1.6 & 1.98 & 8.42 & -0.12 & -0.93 & -1.31 \\
 8S \tablenotemark{d}&  Well Mixed &  4.0 & 2.18 & 8.50 & -0.22 & -1.04 & -1.43 \\
&  & & & & & & \\
Obs. &  & &  1.98 $\pm$ 0.20 & 8.42 $\pm$ 0.06 & -0.31 $\pm$ 0.12 & -1.37 $\pm$ 0.18 & -1.41 $
\pm$ 0.10 \\
\enddata
\tablenotetext{a}{Extra $1.2 \times 108$ \msun \ of primordial gas accreted at 13.0 Gyr. }
\tablenotetext{b}{All the  material expelled by Type II and Ia SNe during the last 
2.5 Gyr is ejected to the IGM. }
\tablenotetext{c}{The upper mass limit for the IMF amounts to 47 \msun.
Fraction of binary systems that become SNIa amounts to 0.02.}
\tablenotetext{d}{The upper mass limit for the IMF amounts to 60 \msun.
Fraction of binary systems that become SNIa amounts to 0.02.}
\end{deluxetable}

In Model 7S
we have reduced  $m_{up}$  from 80 \msun \ to 47 \msun \
and  consequently the predicted O/H value is reduced by 0.22 dex.
This fact produces an increase of the C/O and N/O  predicted values to
-0.12 dex and -0.93 dex, respectively, more than 1.5 $\sigma$ away from the observed values
(see Table 5).
The Fe/O value is adjusted because we have assumed a lower
fraction of binary systems that become SNIa ($Abin=0.02$).
This model does not reproduce the C/O and N/O values because with this IMF there are more LIMS
that are the main producers of  C and N, at these metallicities.
For Model 7S the values of
 $M_{sub}$,  $M_{stars}$, and $M_{rem}$,
 amount to $3.50 \times 10^7$ \msun, $1.72 \times 10^8$ \msun,
and $2.30 \times 10^7$ \msun, respectively.

\begin{figure}[ht]
\plotone{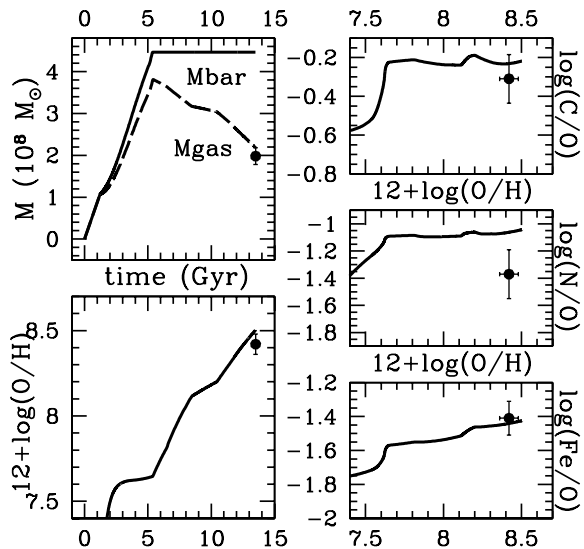}
\caption{
 Model 8S.
Mass formation history and predicted chemical evolution.
This model assumes the gas mass assembly history of small infall model
and includes a well mixed outflow during the initial
4.0 Gyr of the star formation.
The IMF $m_{up}$ is set to 60 \msun \ and the fraction of binary systems that becomes SNIa
is 2 \%.
}
\end{figure}

No previous S model reproduces all the observational constraints.
In particular, those models that match $M_{gas}$, predict higher O/H values
than observed. The extra assumptions included in the models
produce a rise in the  C/O  value, higher than observed.
Consequently, we have computed Model 8S in order
to reproduce   $M_{gas}$,  O/H, and C/O.

One way to reduce the O/H and C/H values, maintaining the C/O ratio, 
is with a  well mixed outflow of longer duration.
In Model 8S, we assume a well mixed outflow during the first 4 Gyr of star formation
and the $m_{up}$ of IMF and the $Abin$ of Model 7L,
the best of the L series models.
Model 8S  reproduces marginally all observational constraints
with the exception of N/O( see Table 5 and Figure 12).

The C budget for  Model 8S is practically the same than for Model 7L (see Table 4)
because both models have the same $SFR$, IMF, $Abin$ and similar O/H.

During the wind phase the outflow rate of Model 8S is equal to 3.13 times the
$SFR$.
The gas mass lost is 7.2 $\times 10^7$ \msun,
3.5 times the mass of stars formed  during the first 4.0 Gyr of the star formation.
Martin (1999)
estimates outflow rates from 1.7 to 4.8 times the $SFR$ for nearby irregular
galaxies with the smallest maximum H I rotation speed of her sample. While
these values are similar to those required by Model 8S, they are smaller
than those required by Model 7L. The observed outflow rates for outbursting
nearby irregular galaxies seem to support Model 8S over Model 7L, but
probably at the time of the formation of the galaxy the accretion and
merging of smaller structures also propitiated larger outflow rates, it
is beyond the scope of this paper to carry this discussion further.

\subsection{The N/O ratio}

All our models fail to fit the N/O ratio,
this could be due to problems with observations and with the adopted N yields.
In what follows we will discuss this problem.

\citet{car05} obtained  an excellent agreement between the present time N/O value predicted
by the solar vicinity model and the N/O values derived from observations of H II regions
of the solar vicinity.
Alternatively,
the N/O values predicted by all our models are higher than those observed
in NGC 6822 by \citet{pei05}.

Moreover the  N/O values derived
from metal poor stars are in agreement with the model by \citet{car05} but not for
metal rich stars, these authors suggest that the disagreement is due to
the N yields adopted in the chemical evolution models.

We consider the M92 yields for C and O for $Z_\odot$ as the best in the
literature and we have used them in this paper. Unfortunately M92 did
not compute the N yields.  In this paper we have adopted the N yields
for $Z_\odot$ by MM02.  The N yield is not self consistent with the C
and O yields. Furthermore since the N abundance is considerably smaller
than that of C and O, the predictions for these two elements are robust,
but the N results for massive stars should be taken with caution.

Romano, Tosi, \& Matteucci (2005) using N yields for LIMS by van der Hoek \& Groenewegen (1997), 
different to those adopted by us, cannot reproduce the N/O values of NGC 1705,
a late type dwarf galaxy. On the other hand, Gavil\'an, Moll\'a, \& Buell (2005) 
using their own N yields are able to match the present-day N/O value of the solar vicinity,
but not the N/O evolution. This problem should be studied further.  

A possible explanation for the low N/O values in H II  regions of NGC 6822
could be due to the escape of an important fraction of the ionizing photons.
If such were the case, the ionization correction factor ($ICF$) for N becomes considerably larger
than the usual $ICF$ adopted \citep{pei05,rel02}.

\section{Discussion} 

The chemical evolution of NGC 6822 depends on $M_{total}$, the MAH, the
baryonic infall as a function of time, and the importance of outflows.
In what follows we will discuss these parameters and how they affect the
different models that we have computed.

\subsection{Total mass}

{From} the cosmological model adopted we have found an
$M_{total} = 2.6 \times 10^{10}$ \msun. In what follows we will
discuss possible upper and lower limits for $M_{total}$. In Figure
13 we present $f_{acc}$ as a function of time for the
minimum, adopted, and maximum $M_{total}$ values.

\begin{figure}[ht]
\plotone{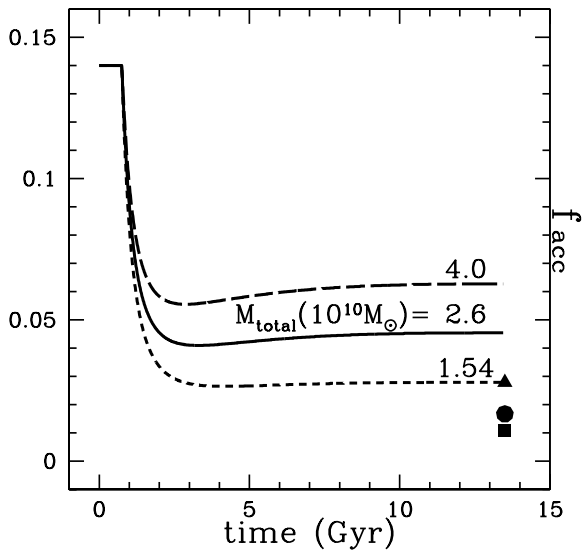}
\caption{
$f_{acc}$, the ratio of the baryonic mass accreted
by galaxy to the total mass, as a function of time for the
minimum, the adopted, and the high  $M_{total}$ values. The
triangle, the circle and the square represent the ratio of
the baryonic mass observed to the $M_{total}$ for the $1.54 \times 10^{10}$ \msun,
$2.6 \times 10^{10}$ \msun, and $4.0 \times 10^{10}$ \msun, respectively.
}
\end{figure}

The minimum $M_{total}$ for the L models is obtained when there are no outflows
present. For the mean MAH and without outflows
we obtain
$M_{total} = 1.54 \times 10^{10}$ \msun, for a smaller mass
the cosmological context predicts a smaller accreted mass than the baryonic
mass observed. We have computed a chemical evolution model
for this total mass and find that 12 + log O/H = 8.66 and C/O = - 0.29 dex.
The O/H value is considerably higher than the observed one.
To fit the observed O/H value we need an O-rich outflow or to reduce
$m_{up}$. In both cases the C/O value becomes considerably
larger than observed. For example a model with $m_{up}$ of 45 \msun \
gives 12 + log O/H = 8.42 in agreement with observations, but it also
gives a C/O = -0.13 dex, 1.5 $\sigma$ higher than the observed value.
Therefore the minimum $M_{total}$ model can be discarded by the
chemical evolution models.
For the S models the minimum $M_{total}$ is smaller than $2.6 \times 10^{10}$ \msun \
but larger than $1.54 \times 10^{10}$ \msun.

The upper limit to $M_{total}$ considered by us is $4 \times 10^{10}$ \msun.
The mass of this model is 1.54 times higher than the mass derived 
in section 3.2.
For this model the accreted gaseous mass amounts to
$25.1 \times 10^{8}$ \msun \ and the ejected mass to $20.8 \times 10^{8}$ \msun.
The ejected gaseous mass is about a factor of three larger than that of Model 7L
and implies that the efficiency and the duration of the outflows have
to be considerably larger than for Model 7L.
If we assume a wind phase of 5.1 Gyr duration, like that of Model 7L,
the outflow phase required amounts to 20.1 times the $SFR$.
This value is about an order of magnitude larger than that derived by Martin (1999)
for nearby irregular galaxies.
 The star formation history,
that drives the outflow, is observationally fixed,
and does not change with the increase of the adopted $M_{total}$.
Moreover the larger
$M_{total}$ provides a larger gravitational field that reduces the
possibility of extended outflows. From these arguments we consider the
adopted upper limit for $M_{total}$ a generous one.

Based on the models by van den Bosch (2003), that predict the virial
masses of disk galaxies, we can also derive the masses for NGC 6822
for the cases with and without outflows. We have two observational
constraints to derive $M_{total}$, the maximum rotational velocity,
that amounts to 55 km s$^{-1}$, and the $K$ luminosity that amounts to
log$(L_{tot}/L_\odot) = 8.5$. The $(L_{tot}/L_\odot)_K$ ratio is
obtained from the NGC 6822 and the solar $M_K$ magnitudes that
amount to -17.91 and 3.33 respectively (see Table 2, Campins et al. 1985,
and Hayes 1985). From the two observational constraints and the
models by van den Bosch with $Z = 0.007$ and no outflows we obtain
$M_{total} \sim 1 \times 10^{10}$ \msun, while for the models with
$Z = 0.007$ and outflow we obtain $M_{total} = (2.0 \pm 0.5) \times 10^{10}$ \msun;
these values are somewhat smaller but in good agreement with those derived by us.
Valenzuela et al. (2005) find $M_{total}$
 $1.93 \times 10^{10}$ \msun \ and $3.4 \times 10^{10}$ \msun \
 for their two favored models of NGC 6822
in good agreement with with our determination.

{From} the computed chemical evolution models we consider that the $M_{total}$
adopted by us is reasonable.  
But if $M_{total}$ were very different from
our adopted value, new chemical evolution models must be computed.

\subsection{Mass assembly history}

In Figure 1 we have presented three MAHs, the slow, the mean, and the fast.
In the models presented in this paper we have adopted the mean MAH. We will
discuss why the other MAHs were not considered. In Figure 14 we show the
$f_{acc}$ for the L  models
as a function of time for the three MAHs under discussion and
under the assumption that $M_{total} = 2.6 \times 10^{10}$ \msun. Due
to the effects of reionization the amount of gas accreted by the galaxy
reaches $8.5 \times 10^{8}$ \msun, $11.7\times 10^{8}$ \msun, and
$17.0 \times 10^{8}$ \msun \ for the slow, the mean, and the fast
MAHs respectively. 
The difference between
the accreted mass and the observed baryonic mass, that amounts to
$4.3 \times 10^{8}$ \msun,
has to be ejected by outflows.

The infall of gas produced by the slow MAH at early times
is not enough to form
the older stars observed in NGC 6822. 
Therefore, based on our models, we  discard the slow MAH.

\begin{figure}[ht]
\plotone{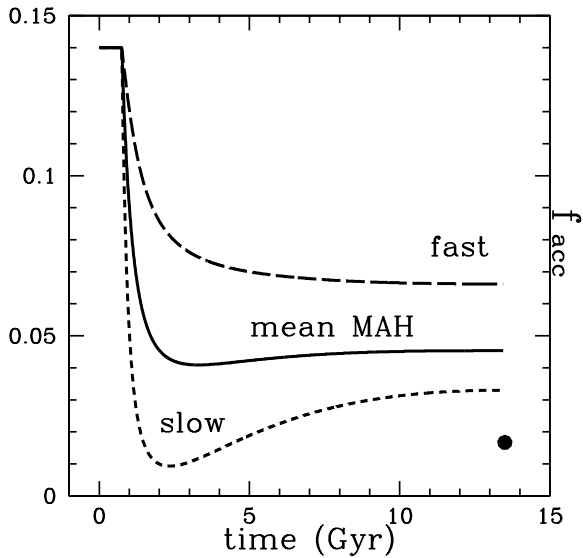}
\caption{
$f_{acc}$, the ratio of the baryonic mass accreted by
the galaxy to the total mass, as a function of time for the slow,
mean, and fast MAHs for $M_{total} = 2.6 \times 10^{10}$ \msun.
The filled circle is the ratio of the baryonic mass observed to the total
mass for the three MAHs.
}
\end{figure}

{From} Figures 1 and 14 it follows that we require to lose a larger
amount of mass due to outflows for the fast than for the mean MAH models.
Therefore the fast MAH models would require an early outflow of longer duration
than the mean MAH models, typically of about 8 Gyr. Furthermore the gravitational
field becomes larger at earlier times for the fast MAH than for the mean MAH models
making it harder for outflows to occur, and the relative large initial gas
content combined with the observed star formation history also prevent the
presence of large outflow rates. These considerations made us discard the fast
MAH models. 

\subsection{Large and small infall models}

As mentioned in section 3.4 several authors 
have suggested that a significant fraction of baryons
is not accreted by the galaxies due to large-scale shock heating
of the gas. To consider
this possibility we have computed a series of S models
assuming that the evolution up to $t = 5.37$ Gyr is the same as for the
L models, but after this time accretion stops for the 
S models. The  S models accrete only 60\% of
the mass accreted by the L models (see Figures 2 and 3).

The total mass accreted by the S models amounts to
$7 \times 10^{8}$ \msun, and since the baryonic mass at present is
$4.3 \times 10^{8}$ \msun, these models only require to lose
$2.7 \times 10^{8}$ \msun. In Figure 15 we present $f_{gal}$, the
ratio of the mass in baryons that remains in the galaxy to the
total mass as a function of time.

\begin{figure}[ht]
\plotone{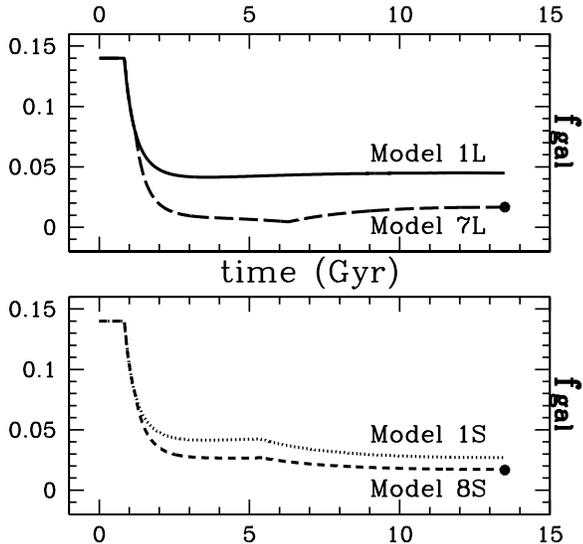}
\caption{
$f_{gal}$, the ratio of the mass in baryons that remains
in the galaxy to the total mass, as a function of time. In the upper
panel we show two large infall models: Model 1L,
without outflows, and Model 7L our best model that includes an early outflow.
In the lower panel we show two small infall models: Model 1S with no outflows
and Model 8S that includes an early outflow. Note that for Models
1S and 1L $f_{gal} \equiv f_{acc}$.
Filled circles represent the observational data.
}
\end{figure} 

In general the S models predict higher enrichments
in O/H and C/H than the L models (see Tables 3 and 5). 
Model 8S,
the best model of the S series, requires a well mixed outflow with
a duration of 4.0 Gyr at the beginning of the $SFR$; 
its O/H, C/H, and $M_{gas}$ values are
1.25$\sigma$, $1.4\sigma$ and $1\sigma$ higher than observed. 
Based on
the chemical evolution models we favor the large-infall series over the
small-infall one, particularly we prefer Model 7L over Model 8S.

If the effect of large-scale shock heating on gas accretion involves a fraction
of baryonic matter smaller than that taken into account by the S series models,
it will be possible to build successful galactic chemical evolution models,
intermediate between Models 7L and 8S.

\section{Conclusions}

We have derived the $SFR$ fort NGC 6822 based on observations and photometric evolution models.
Our photometric models based on the adopted $SFR$,
the Salpeter IMF, the KTG IMFs, and three $Z(t)$ functions 
(those predicted by our best chemical evolution models, 7L and 8S,
and that by Wyder)
predict absolute magnitudes and colors that
agree with the observed ones at the one $\sigma$ level.
This result implies that our derived $SFR$ is robust and can be used
for different applications related to NGC 6822.

Based on the adopted $SFR$ we present 15 chemical evolution models for NGC 6822.
All models evolve according to the mass assembly history of non baryonic mass predicted by
a cosmological context.
To distinguish among them we have used five additional observational constraints:
$M_{gas}$, O/H, C/O, N/O, and Fe/O.

{From} the cosmological context adopted and our chemical and photometric evolution models
for NGC 6822 we find that 
$1.54 < M_{total}(10^{10} \msun) < 4.0 $.
The adopted mass for the 15 models presented in Tables 3 and 5 is 
$2.6 \times 10^{10}$ \msun.
Based on similar arguments, of the three MAHs considered,
 we adopt the mean MAH for all our models.

For the  models of the large infall series,
 we have  assumed that during accretion the universal baryon fraction
is only reduced by reionization. 
For the  models of the small infall series,
 we have assumed that accretion of baryonic matter is
further reduced by large-scale gas shock heating. 
The best models of each series,
Models 7L and  8S, produce a reasonable fit to the observational constraints,
but Model 7L is somewhat better.

Models  without well mixed outflows are not successful in reproducing
the observational constraints because the amount of baryonic gas accreted by
the galaxy is higher than the amount of baryonic matter derived from
observations.
Moreover based on our adopted $SFR$ the thermal heating produced by SNe, 
under the assumption of no outflows, 
is considerably larger than the binding energy of the observed gas content. 
These two results imply that
the presence of  outflows of well mixed material is needed.
We argue that these outflows most likely occurred early on in the history of the galaxy.

In order to comply with the presence
of an early outflow,
and the observed O/H and C/H values, we have computed our best models (Model 7L and 8S)
that have a KTG IMF with $m_{up}$ of 60 \msun \ while for the  solar vicinity
$m_{up}$ amounts to 80 \msun.
That is the number of SN of Type II per generation
of stars has to be smaller in NGC 6822 than in the solar vicinity.
A lower value of $m_{up}$ for NGC 6822 than for the solar vicinity
is in agreement with recent results on dwarf galaxies by others authors.

According to Models 7L and 8S and the observed Fe/O value
the fraction of binary systems that become SNIa in NGC 6822
is 2 \% while  in the solar vicinity amounts to 5 \%.

The fractions of C produced by MS ($m >$ 8\msun), LIMS ($m <$ 8\msun),
and SNIa amount to 60 -- 63\%,  38 -- 36\%, and 1 --2\% respectively. 
On the other hand, in the solar vicinity, these values amount to
48\%, 50\%, and 2\%, respectively.

The N/O values predicted by all our models are higher than observed,
the discrepancy could be due  to
the N ionization correction factor adopted to derive the N abundance in H II regions
and to the N stellar yields adopted by our chemical evolution models.

The results obtained in this paper apply to an isolated typical irregular galaxy,
but do not apply to irregular galaxies that have suffered strong tidal effects.

Further progress in modelling NGC 6822 can come if a more precise $M_{total}$ is
determined, and 
if the issue 
about the large-scale gas shock heating is finally settled.

\vskip 1cm


L.C. thanks Gustavo Bruzual for instructing her
how to use the GALAXEV code including a metal enrichment history,
and for computing a set of photometric properties for instantaneous-burst models
based the KTG IMF.
L.C. is also grateful to Ted K. Wyder for several illuminating explanations about
the star formation history of NGC 6822. 
P.C. thanks A. Kravtsov for kindly providing 
information about mass assembly histories in simulations.
We acknowledge Octavio Valenzuela for valuable conversations and 
thank Vladimir Avila-Reese  for helpful suggestions.
We are grateful to Donatella Romano for a critical reading of the manuscript.
We also acknowledge to the anonymous referee for a careful reading of the manuscript
and many excellent suggestions.
LC's work is supported by CONACyT grant 36904-E.
PC acknowledges support by CONACyT grant 36584-E.
MP received partial support from DGAPA UNAM (grant IN114601).

\end{document}